\documentclass[11pt]{article}
\usepackage{psfrag}
\usepackage{graphicx,epsfig}
\newcommand{\be}{\begin{equation}}
\newcommand{\en}{\end{equation}}
\newcommand{\bea}{\begin{eqnarray}}
\newcommand{\ena}{\end{eqnarray}}
\newcommand{\lbl}[1]{\label{eq:#1}}
\newcommand{\rf}[1]{(\ref{eq:#1})}
\newcommand{\lblfig}[1]{\label{fig:#1}}
\newcommand{\fig}[1]{\ref{fig:#1}}
\newcommand{\lbltab}[1]{\label{tab:#1}}
\newcommand{\Table}[1]{\ref{tab:#1}}

\def\today{\ifcase \month\or
  January\or February\or March\or April\or May\or June\or
    July\or August\or September\or October\or November\or December\fi
      \space\number\day,\space
        \number\year }
\newcommand{\lapprox}{%
\mathrel{%
\setbox0=\hbox{$<$}\raise0.6ex\copy0\kern-\wd0\lower0.65ex\hbox{$\sim$}}}

\newcommand{\gapprox}{%
\mathrel{%
\setbox0=\hbox{$>$}\raise0.6ex\copy0\kern-\wd0\lower0.65ex\hbox{$\sim$}}}

\newcommand{\mpi}{m_\pi}
\newcommand{\mk} {m_K}
\newcommand{\mpid}{m^2_\pi}

\newcommand{\mkd} {m^2_K}

\newcommand{\mpl}{m_+}

\newcommand{\mpd}{{m^2_+}}
\newcommand{\mmd}{{m^2_-}}

\newcommand{\ssd}{s'_\Delta}
\newcommand{\ttd}{t'_\Delta}
\newcommand{\sd}{s_\Delta}
\newcommand{\ud}{u_\Delta}

\newcommand{\Kbar}{\overline{K}}
\newcommand{\undemi}{{1\over2}}
\newcommand{\tm}{{t_{m}} }

\def\im{ {\rm Im}\,}
\def\re{ {\rm Re}\,}
\def\upi{ {1\over\pi}}
\def\udemi{ {1\over2}}
\def\tdemi{ {3\over2}}
\def\utier{ {1\over3}}
\def\dtier{ {2\over3}}
\def\qtier{ {4\over3}}
\def\intpik{\int_{m^2_+}^\infty}
\def\intpipi{\int_{4m_\pi^2}^\infty}
\setbox4=\hbox{$-$}
\def\pvint#1#2{\copy4\kern-1.1\wd4\hbox{$\displaystyle\int_{#1}^{#2}$}}
\def\xlss{\lambda_{s'}}
\def\xlssb{\lambda^b_{s'}}
\def\xls{\lambda_{s}}
\def\Deld{\Delta^2}
\def\Sig{\Sigma}

\def\ar{{\alpha^\rho}}
\def\ak{{\alpha^{K^*}} }

\begin{document}
\begin{titlepage}
\begin{flushright}
HISKP-TH-03/18\\
IPNO/DR-03-08/\\
LPT-ORSAY/03-76\\
\today
\end{flushright}
\begin{center}
{\Large\bf  A new analysis of $\pi K$ scattering\\
from Roy and Steiner type equations}
\footnote{Work supported in part by the EU RTN contract 
HPRN-CT-2002-00311 (EURIDICE) and by IFCPAR contract 2504-1.}\\[1.5cm]

{\large P. B\"uttiker$^a$, S. Descotes-Genon$^b$ and B. Moussallam$^c$} 

{\sl$^a$\ Helmholtz-Institut f\"ur Strahlen- und Kernphysik,}\\
{\sl Universit\"at  Bonn, D-53115 Bonn, Germany}

{\sl$^b$\ Laboratoire de Physique Th\'eorique\footnote{
LPT is an unit\'e mixte de recherche du CNRS et de l'Universit\'e Paris-Sud
(UMR 8627).} }\\
{\sl Universit\'e Paris-Sud, F-91406 Orsay, France}

{\sl$^c$\ Institut de Physique Nucl\'eaire\footnote{
IPN is an unit\'e mixte de recherche du CNRS et de l'Universit\'e Paris-Sud
(UMR 8608).} }\\ 
{\sl Universit\'e Paris-Sud, F-91406 Orsay, France}
\end{center}
\vfill

\begin{abstract}
With the aim of generating new constraints on the OZI suppressed couplings 
of chiral perturbation theory a set of six equations of the
Roy and Steiner type for the $S$- and $P$-waves of 
the $\pi K$ scattering amplitudes is derived. The range of validity and 
the multiplicity of the solutions are discussed. 
Precise numerical solutions are obtained in the range $E\lapprox 1$ GeV which
make use as input, for the first time, 
of the most accurate experimental data available
at $E\gapprox 1$ GeV for both $\pi K\to\pi K$ and $\pi\pi\to K\overline{K}$
amplitudes. Our main result is the determination of a narrow  allowed region
for the two S-wave scattering lengths.
Present experimental data below 1 GeV are found
to be in generally poor agreement with our results. 
A set of threshold expansion parameters, as well as sub-threshold
parameters are computed. For the latter, a matching with the $SU(3)$
chiral expansion at NLO is performed.
\end{abstract}
\vfill

\end{titlepage}
\section{Introduction}

Scattering amplitudes of pseudo-Goldstone bosons at low energies
probe with a unique sensitivity the scalar-source sector of Chiral
Perturbation Theory (ChPT)~\cite{gl84,gl85}. 
For instance, recent progress in the domain 
of $\pi\pi$ scattering has provided  valuable information on
the $SU(2)$ chiral limit where the masses of the $u,d$ quarks are
set to zero. 
For this purpose, the $\pi\pi$ Roy equations, which have been
extensively studied in the past~\cite{basdevant,bfp,pennington},  
were re-analyzed~\cite{acgl} (in particular, a formulation as a boundary
value problem was developped)
and solved numerically~\cite{acgl,cgl} (see also~\cite{kll}). These equations
constrain the low-energy $\pi\pi$-scattering amplitude by exploiting
simultaneously theoretical requirements and data at higher
energies. New data on $Kl_4$ decays from the E865 
experiment~\cite{pislak} could thus be
studied with the help of the solutions to the Roy equations and 
a bound on the coupling constant $\bar\ell_3$ 
of the $SU(2)$ chiral Lagrangian~\cite{cgl2} was derived for the first time. 
Constraints on the $SU(2)$ quark condensate were  also
obtained along similar lines~\cite{dfgs}.

In a parallel way, scattering amplitudes involving both pions and kaons
at very low energy should allow one to unveil features of the $SU(3)$
chiral vacuum, i.e. that in the limit where $m_u$, $m_d$ and $m_s$ 
vanish. The structure of the $SU(3)$ chiral vacuum is  worth studying 
for its own sake, since $SU(3)$ ChPT provides 
relations between many low-energy 
processes involving $\pi$-, $K$- and $\eta$-mesons. In addition,
it is interesting to compare $SU(2)$ and $SU(3)$ chiral limits,
especially in the scalar sector. A sizable difference between the
two limits would indicate that sea-quark effects are particularly
significant in the case of the strange quark~\cite{dgs1,dgs2}.
In previous works~\cite{ab,abm}  it was shown that the
ratio of the pion's decay constant in $SU(2)$ and $SU(3)$ chiral limits
could be determined from a sum rule based on $\pi K$  scattering
amplitudes. The deviation of this ratio from 1 would indicate a violation
of the large-$N_c$ approximation. Let us emphasize that the latter 
is often relied upon to attribute values to some 
$O(p^4)$ couplings arising in the scalar sector of the chiral 
Lagrangian~\cite{gl85,amoros}. Our work is motivated by the desire of
determining from $\pi K$-scattering experimental data as many chiral
couplings as possible (in principle, five out of the ten 
independent $O(p^4)$ couplings of the $SU(3)$ chiral 
Lagrangian~\cite{bkm1,bkm2}),
without relying on the large-$N_c$ approximation. In this paper, we 
provide the first step of this analysis, by deriving the analogue
of Roy equations for the $\pi K$ system and solving them numerically.
A simple matching with the $SU(3)$ expansion is performed while more
detailed comparisons with $SU(2)$ and $SU(3)$ expansions are left for 
future work.
Further motivation for the study of $\pi K$ scattering can be found in 
refs.~\cite{jop,jop1}.

In the case of $\pi\pi$ scattering, 
Roy observed~\cite{roy} that general properties of 
analyticity, unitarity, combined with 
crossing symmetry, lead to  a set
of non-linear integral equations that the $S$- and $P$-partial waves
must satisfy. A similar program was carried out by Steiner for
$\pi N$ scattering~\cite{steiner}.
Given experimental input at high energies (typically
$E\gapprox 1$ GeV), Roy-Steiner (RS) equations constrain 
the low-energy behaviour of partial-wave amplitudes.
In the present paper, we derive and perform a detailed analysis of a system of
RS equations for $\pi K$ scattering. 
In this case, $s-t$ crossing relates 
the $\pi K\to \pi K$ and the $\pi\pi\to K\overline{K}$
amplitudes, leading to six coupled equations that
involve the four  $\pi K$ $S$ and $P$  partial-wave amplitudes
$f_0^{1/2},\ f_1^{1/2},\ f_0^{3/2},\ f_1^{3/2} $ and the two  
$\pi\pi\to K\overline{K}$ amplitudes $g_0^0,\ g_1^1$. Equations
of a similar kind have
been considered earlier~\cite{bonnier,lang,johannesson}. However, some
approximations were invoked in the treatment of these equations and,
moreover, no accurate experimental input data were available at that time. 
Since then, high-statistics production experiments have
been performed for both $\pi K\to \pi K$~\cite{estabrooks,aston} 
and $\pi\pi\to K\overline{K}$ amplitudes~\cite{cohen,etkin}. These 
experiments provide the necessary input data for the RS equations
with a level of accuracy comparable to the case of $\pi\pi$ scattering. 
Experimental data at lower energies should also be available in the near
future: the FOCUS experiment~\cite{focus} 
has demonstrated the feasibility of determining the $\pi K$ 
$S$-wave phase shifts
at energies lower that 1 GeV from the weak decays of D mesons~\cite{pham},
P-wave phase shifts should be measured soon in
$\tau$ decays~\cite{weinstein} and finally, 
direct determinations of combinations of 
$S$-wave scattering lengths are expected from planned 
experiments on kaonic atoms~\cite{katom}.

The plan of the paper is as follows. After reviewing the notation, we derive
the set of RS equations that we intend to solve. The setting is similar to
a previous  work~\cite{johannesson} but we differ in the number of
subtractions used in the dispersive representations.
We aim here at an optimal use of the energy region where accurate 
experimental data are available, while avoiding to
rely on slowly convergent sum rules.
After discussing the domains of validity of such equations,
we explain our treatment of the available experimental input and of the 
asymptotic regions. Next, we start solving  the equations. One first
eliminates $g_0^0$ and $g_1^1$ and the remaining four equations then
have a similar structure to the $\pi\pi$ Roy equations 
such that recent results 
concerning the multiplicity of the solutions~\cite{gasser-wanders,wanders} 
can be exploited. Finally, we turn to the numerical resolution and 
discuss the resulting constraints on the S-wave scattering lengths. 
Finally, the $\pi K$ amplitudes near and below threshold are constructed 
and estimates for the $O(p^4)$ chiral coupling constants obtained from
matching with the $SU(3)$ expansion are given.

\section{Derivation of the equations} \label{sec:ampldisp}

\subsection{Notation}

Let us recall briefly some standard notation~\cite{lang-rev}. 
Firstly, we define from the pion and kaon masses
\be\lbl{massdef}
m_\pm=\mk\pm \mpi,\ \Sigma=\mkd+\mpid,\ \Delta=\mkd-\mpid\ .
\en
In this paper, exact isopin symmetry will always be  assumed.
In the isospin limit, there are two independent
$\pi K$ amplitudes $F^I(s,t)$, with isospin $I={1\over2}$ and $I={3\over2}$. 
Making use of $s-u$ crossing,
the $I={1\over2}$ amplitude can be expressed in terms of the $I={3\over2}$ one,
\be
F^{1\over2}(s,t,u)=-{1\over2}F^{3\over2}(s,t,u)+
                    {3\over2}F^{3\over2}(u,t,s)\ .
\en
It is convenient to introduce the amplitudes $F^+$ and 
$F^-$ which are, respectively, even and odd under $s-u$ crossing.  
In terms of isospin amplitudes, they are defined as
\bea
&&F^+(s,t,u)={1\over3} F^{1\over2}(s,t,u)
+            {2\over3} F^{3\over2}(s,t,u)\nonumber\\
&&F^-(s,t,u)={1\over3} F^{1\over2}(s,t,u)
-            {1\over3} F^{3\over2}(s,t,u)\ .
\ena

The partial-wave expansion of the $\pi K$ isospin amplitudes is defined
as
\be\lbl{Fpw}
F^I(s,t)=16\pi\sum_l (2 l+1) P_l(z_s) f^I_l(s)\ .
\en
where $P_l(z)$ are the standard Legendre polynomials
and  $z_s$ is the cosine of the $s$-channel scattering angle 
\be
z_s=1+{2st\over\lambda_s}
\quad {\rm with}\ \lambda_s=(s-m^2_+)(s-m^2_-) \ .
\en
In a similar way we can expand $F^+$ and $F^-$, and the corresponding 
partial-wave projections are denoted by $f^+_l(s)$ and $f^-_l(s)$. 
The amplitudes can be projected over the partial waves through
\begin{equation} \lbl{projects}
f^I_l(s)=\frac{s}{16\pi \lambda_s} \int_{-\lambda_s/s}^0 dt
  \ P_l(z_s) F^I(s,t)\ .
\end{equation}
The values of the amplitudes at threshold define the $S$-wave 
scattering lengths, with the following conventional normalization
\be
 a_0^I ={2\over\mpl} f_0^I(\mpd)
\en
(and  similarly for $a_0^\pm$ in terms of $f_0^\pm(\mpd)$). 

Under $s-t$ crossing, one generates the $I=0$ and $I=1$ 
$\pi\pi\to K\overline K$
amplitudes,
\bea
&&G^0(t,s,u)=\sqrt6 F^+(s,t,u)\nonumber\\
&&G^1(t,s,u)=2      F^-(s,t,u)\ .
\ena
The partial-wave expansion of the $\pi\pi\to K\overline K$ amplitudes 
is conventionally defined as
\be\lbl{Gpw}
G^I(t,s)=16\pi\sqrt2\,\sum_l (2 l+1) [q_\pi(t) q_K(t)]^l P_l(z_t) 
g^{I}_l(t)\ ,
\en
where the summation runs over even (odd) values of $l$ 
for $I=0$ ($I=1$) due to Bose symmetry in the $\pi\pi$ channel. 
In this expression the momenta $q_\pi$, $q_K$ and
the cosine of the $t$-channel scattering angle $z_t$ are given by
\be
q_P(t)={1\over 2}\sqrt{t-4m^2_P},\quad
z_t={s-u\over 4 q_\pi(t)q_K(t)}\ .
\en

The relations between these partial-wave amplitudes and 
the $S$-matrix elements are easily worked out
\bea
&&\left[ S^I_l(s)\right]_{\pi K\to\pi K}  
=1+2i{\sqrt{\lambda_s}\over s} \theta(s-\mpd) f^I_l(s)
\nonumber\\
&&\left[ S^I_l(t)\right]_{\pi\pi\to K\overline K} =4i
{(q_\pi(t) q_K(t))^{l+1/2}\over\sqrt{t}} \theta(t-4\mkd) g^I_l(t)\ .
\ena

\subsection{Fixed-$t$ based dispersive representation}

To derive RS equations,
we assume the validity of the Mandelstam double-spectral
representation~\cite{mandelstam0} from which one can derive
a variety of dispersion relations (DR's) for
one variable~\footnote{For the $\pi K$ amplitude, the existence of 
fixed-$t$ DR can be established on more general grounds in a finite
domain of $t$~\cite{mandelstam,fix-t-valid}.}. According to the Froissart 
bound~\cite{froissart}, two
subtractions are needed at most for $F^+$ and one subtraction for $F^-$
(because $s-u$ can be factored out in the latter case). 
More detailed information about asymptotic
behaviour is provided by Regge 
phenomenology~\cite{regge-review}, according to
which two subtractions are indeed necessary for $F^+$ 
while an unsubtracted DR is expected to converge for $F^-$.
However, convergence is rather slow in the latter case 
since the integrand behaves like $(s')^{-3/2}$ asymptotically.
Therefore, we choose to make use of a once-subtracted
DR for $F^-$ in order to improve the
convergence and reduce the sensitivity to the high-energy domain. 

Fixed-$t$ DR's for $F^+$ and $F^-$, with the number of subtractions
as discussed above can be written in the following form
\bea\lbl{fixtdr}
&& F^+(s,t)=c^+(t)+{1\over\pi}\int_\mpd^\infty ds'\,\left[
{1\over s'-s}+{1\over s'-u}-{2s'-2\Sigma-t\over \xlss}
\right] \im F^+(s',t)\ .
\nonumber\\
&& {F^-(s,t)\over s-u}=c^-(t)+
{1\over\pi}\int_\mpd^\infty ds'\,\left[{1\over(s'-s)(s'-u)}
-{1\over\xlss}\right] \im F^-(s',t)\ .
\ena
These expressions involve two unknown functions of $t$: $c^+(t)$ and
$c^-(t)$. The basic idea for determining these functions is to invoke
crossing~\cite{roy,steiner}, which can be implemented in various ways:
for instance, one can use fixed-$s$ or fixed-$(s-u)$ DR's. 
After trying several possibilities, we found that DR's
at fixed $us$  provide the largest domain of applicability
(these relations, sometimes called hyperbolic DR's, were exploited
in ref.~\cite{johannesson}). We start with
a special set of hyperbolic DR's (more general hyperbolic DR's  
will be considered later) in which 
\be
us=\Delta^2\ . 
\en
The condition above fixes $s$ and $u$ to be functions of $t$
\begin{eqnarray}
s&\equiv& s_\Delta(t) ={1\over2}\left(2\Sigma-t+ 
\sqrt{(t-4\mpid)(t-4\mkd)}\right)\ 
\nonumber\\
u&\equiv& u_\Delta(t) ={1\over2}\left(2\Sigma-t- 
\sqrt{(t-4\mpid)(t-4\mkd)}\right)\ .
\end{eqnarray}

According to Regge theory, 
the function $F^+(s_\Delta,t)$ satisfies a once-subtracted
DR which is slowly converging. Like in the case of the fixed-$t$ DR for 
$F^-$, we choose to improve the convergence by using
a twice-subtracted representation. On the other hand, the 
function $F^-(s_\Delta,t)$ is expected to satisfy an 
unsubtracted DR which is well converging. 
Making use of the fact that $s_\Delta(0)=\mpd$, these DR's can be written
in the following way
\bea\lbl{hypdr}
&&F^+(\sd,t)=  8\pi m_+ a_0^+ +b^+ t 
+{1\over\pi}\int_\mpd^\infty ds'\,\left[
{2s'-2\Sigma+t\over\xlss+s't} -{2s'-2\Sigma-t\over\xlss}
\right] \im F^+(s',t'_\Delta)
\nonumber\\
&&\phantom{F^+(\sd,t)=  8\pi m_+ a_0^+ +b^+ t }
+{t^2\over\sqrt6\pi}\int_{4\mpid}^\infty {dt'\over (t')^2(t'-t)}
\im G^0(t',\ssd)
\nonumber\\
&&{F^-(\sd,t)\over \sd-\ud}={8\pi m_+ a_0^-\over \mpd-\mmd}+
{1\over\pi}\int_\mpd^\infty ds'\,\left[{1\over \xlss+s't}
-{1\over\xlss}\right] \im F^-(s',\ttd)
\nonumber\\
&&\phantom{{F^-(s,t)\over s-u}={8\pi m_+ a_0^-\over \mpd-\mmd}}
+{t\over2\pi}\int_{4\mpid}^\infty {dt'\over t'(t'-t)} 
\im { G^1(t',\ssd)\over
\sqrt{(t'-4\mpid)(t'-4\mkd)}}\ .
\ena
In these equations, we have used the following notation
\be
\ssd=\sd(t'),\quad \ttd=2\Sigma -s' -{\Delta^2\over s'}\ , 
\en
together with the relation  
$(s'-s_\Delta(t))(s'-u_\Delta(t))=\lambda_{s'}+s't$ .

These representations involve three subtraction constants: the 
two scattering lengths $a^+_0$, $a^-_0$ and an additional
parameter denoted $b^+$. 
Let us now show that the latter can be computed through 
a rapidly convergent sum rule. 
We notice first that $a^-_0$ and $b^+$ satisfy slowly convergent sum rules,
\bea\lbl{slowsr}
&&{8\pi m_+ a_0^-\over \mpd-\mmd}=
{1\over\pi}\int_\mpd^\infty {ds'\over\xlss} \im F^-(s',\ttd)
+{1\over2\pi}\int_{4\mpid}^\infty {dt'\over t'} \im{ G^1(t',\ssd)\over
\sqrt{(t'-4\mpid)(t'-4\mkd)}}\ .
\nonumber\\
&&b^+={-1\over\pi}\int_\mpd^\infty {ds'\over\xlss} \im F^+(s',t'_\Delta)
+{1\over\sqrt6\pi}\int_{4\mpid}^\infty {dt'\over(t')^2} \im G^0( t',\ssd)\ .
\ena
By combining these two sum rules, we can express the parameter
$b^+$ as a sum rule which has better convergence property: 
\bea\lbl{bsumr}
&&b^+= {8\pi m_+ a_0^-\over \mpd-\mmd}
-{1\over\pi}\int_\mpd^\infty {ds'\over\xlss} 
\im\left[ F^+(s',\ttd)+F^-(s',\ttd)\right]
\nonumber\\
&&\phantom{{8\pi m_+ a_0^-\over \mpd-\mmd}-b^+}
+{1\over\pi}\int_{4\mpid}^\infty {dt'\over t'}
\im \left[ {G^0(t',\ssd)\over\sqrt6 t'} -
{G^1(t',\ssd)\over2\sqrt{(t'-4\mpid)(t'-4\mkd)}} 
\right]\ .
\ena
Why does this sum rule converge more quickly ? In the first integral, the 
combination $F^+ +F^- $ appears, which is the amplitude for the process
$\pi^+ K^-\to\pi^+ K^-$. The asymptotic region of the integrand
corresponds to $s\to\infty$, $u\to 0$. The amplitude in this region is 
controlled by the Regge trajectories in the $u-$channel which is exotic,
leading to a fast decrease of the integrand. 
In the second integral, the high-energy tail involves
the combination ${1\over\sqrt6}G^0(t',s') - 
{1\over2}G^1(t',s')$ for $t'\to\infty$ and $s'\to0$. 
The leading Regge contributions
are generated by the $K^{**}$ and $K^*$ trajectories
\be
\lim_{t\to\infty,\ s\to 0}\im \left[ {1\over\sqrt6}G^0(t,s) - 
{1\over2}G^1(t,s) \right]=
\beta_{K^{**}}(s)  t^{\alpha_{K^{**} }(s)} -
\beta_{K^{*}}(s)  t^{\alpha_{K^{*} }(s)} \ .
\en
This difference would vanish if Regge trajectories satisfied
exactly the property of exchange degeneracy.
In nature, this property is not exact but it
has long been observed to be approximately fulfilled 
~\footnote{The underlying reason
for this property is not understood but 
could be related to the possibility that
the large-$N_c$ limit of QCD is described by a 
string theory~\cite{thooft,string}.} 
(see e.g.~\cite{regge-review} ), which should lead to a significant 
suppression of the integrand at high energies. 
Therefore, the 
two integrals involved in eq.~\rf{bsumr} are expected to converge
quickly, providing a determination of $b^+$ with only a mild sensitivity
to high energies.

Combining  the two dispersive representations eqs.~\rf{fixtdr}
and \rf{hypdr} for the amplitudes
$F^+$ and $F^-$, the subtraction functions in eqs.~\rf{fixtdr} 
get determined in terms of the two $S$-wave scattering lengths
and we obtain the following representation for the two amplitudes 
\bea\lbl{fixtfin}
&&F^+(s,t)=8\pi m_+ a_0^+ +b^+ t 
+{1\over\pi}\int_\mpd^\infty ds'\,\left[
{1\over s'-s} +{1\over s'-u} -{2s'-2\Sigma+t\over\xlss+s't}
\right] \im F^+(s',t)
\nonumber\\
&&\phantom{F^+(s,t)=8\pi m_+ a_0^+ +b^+ t}
+{1\over\pi}\int_\mpd^\infty ds'\,\left[
{2s'-2\Sigma+t\over\xlss +s't} -{2s'-2\Sigma-t\over \xlss}
\right] \im F^+(s',\ttd)
\nonumber\\
&&\phantom{F^+(s,t)=8\pi m_+ a_0^+ +b^+ t}
+{t^2\over\sqrt6\pi}\int_{4\mpid}^\infty {dt'\over (t')^2(t'-t)}
\im G^0(t',\ssd)\ ,
\nonumber\\
&&F^-(s,t)={8\pi m_+ a_0^-\over \mpd-\mmd}(s-u)+
{1\over\pi}\int_\mpd^\infty ds'\,\left[{1\over s'-s}-{1\over s'-u}
-{s-u\over\xlss+s't}\right] \im F^-(s',t)
\nonumber\\
&&\phantom{F^-(s,t)={8\pi m_+ a_0^-\over \mpd-\mmd}(s-u)}
+(s-u)\Bigg\{
{1\over\pi}\int_\mpd^\infty ds'\,
\left[{1\over \xlss+s't}-{1\over\xlss}\right]   \im F^-(s',\ttd)
\nonumber\\
&&\phantom{F^-(s,t)={8\pi m_+ a_0^-\over \mpd-\mmd}(s-u)}
+{t\over2\pi}\int_{4\mpid}^\infty {dt'\over t'(t'-t)} 
\im{ G^1(t',\ssd)\over
\sqrt{(t'-4\mpid)(t'-4\mkd)}}
\Bigg\}
\ena
where the parameter $b^+$ is to be expressed in the terms of the sum rule
eq.~\rf{bsumr}.
The domain of applicability of this representation is limited by the
domain of validity of the fixed$-t$ DR's, eq.~\rf{fixtdr}.
In sec.~\ref{sec:validity}, we will show that the fixed-$t$ DR's hold
for $t<4\mpid$, which enables us to perform the projection of eq.~\rf{fixtfin}
on $\pi K\to \pi K$ partial waves. We will also  need a representation which is
valid for $t \ge 4\mpid$ in order to obtain equations
for the $\pi\pi \to K \bar{K}$ partial waves. 
For this purpose, we now consider a family of hyperbolic DR's.

\subsection{Fixed $us$ dispersive representation}

Let us consider a general family of hyperbolic DR's for which
\be
us=b\ 
\en
is fixed. $b$ is a parameter with (a priori) arbitrary values
and should not be confused with the subtraction constant $b^+$ 
introduced in the previous section. 
We write down a twice-subtracted representation
for $F^+(s_b,t)$ and a once-subtracted one for $F^-(s_b,t)$,
\bea\lbl{hypdrb}
&& F^+(s_b,t)=f^+(b)+t h^+(b)+\upi\intpik ds'\left[
{2s'-2\Sig+t\over \xlssb +s't}- {2s'-2\Sig-t\over \xlssb}\right]
\im F^+(s',t'_b)
\nonumber\\
&&\phantom{F^+(s_b,t)=h^+(b)+t f(b)}
+{t^2\over\sqrt6\pi}\intpipi {dt'\over t^{'2}(t'-t)} \im G^0(t',s'_b)\ ,
\nonumber\\
&& {F^-(s_b,t)\over s_b-u_b} =f^-(b)
+\upi\intpik ds' \left[ {1\over\xlssb+s't}
-{1\over\xlssb}\right]\im F^-(s',t'_b)\nonumber\\
&&\phantom{{F^-(s_b,t)\over s_b-u_b} =h^-(b)}
+{t\over2\pi}\intpipi {dt'\over t'(t'-t)} \im {G^1(t',s'_b)\over s'_b-u'_b}
\ena
with the notation
\bea
&& s'_b=\udemi\left(2\Sigma-t'+ \sqrt{ (2\Sigma-t')^2-4b}\right)\nonumber\\
&& t'_b=2\Sigma-s'-{b\over s'}\\
&& \xlssb =(s')^2 - 2\Sig s' +b\nonumber\ .
\ena
The representations eqs.~\rf{hypdrb} are a generalization of the DR's 
eqs.~\rf{hypdr} derived for $us=\Delta^2$. They involve three unknown 
functions of $b$: $f^+(b)$, $f^-(b)$ and $h^+(b)$ (which generalize the 
subtraction constants of eqs.~\rf{hypdr} )
The two functions $f^+(b)$, $f^-(b)$ can be determined by matching 
eqs.~\rf{hypdrb} with the representations eqs.~\rf{fixtfin}
at the point $t=0$ (which lies inside their domain of validity). Next, the
function $h^+(b)$ can be expressed as a rapidly convergent sum rule 
analogous to eq.~\rf{bsumr}. 
Putting things together, one finally obtains the following representations
involving the two $S$-wave scattering lengths $a_0^+$, 
$a_0^-$ as the only arbitrary constants,
\bea\lbl{hypdrbfin}
&& F^+(s_b,t)=8\pi m_+ \left(a_0^+ +t {a_0^-\over \mpd-\mmd}\right)\nonumber\\
&&\qquad \phantom{F^+}
+\upi\intpik ds'\,\Bigg\{ {2s'-2\Sig+t\over\xlssb+s't} \im F^+(s',t'_b)
-{2s'-2\Sig\over\xlssb} \im [ F^+(s',t'_b)-F^+(s',0)]
\nonumber\\
&&\qquad \phantom{F^+ \upi\intpik ds'\Big\{ }
-{t        \over\xlssb} \im [ F^-(s',t'_b)-F^-(s',0)]\nonumber\\
&&\qquad \phantom{F^+ \upi\intpik ds'\Big\{ }
-{2s'-2\Sig\over\xlss} \im F^+(s',0)-{t        \over\xlss} \im F^-(s',0)\Bigg\}
\nonumber\\
&&\qquad \phantom{F^+}
+{t\over\pi}\intpipi {dt'\over t'}\,\left[ {\im G^0(t',s'_b)\over\sqrt6(t'-t)}
-\im {G^1(t',s'_b)\over2(s'_b-u'_b)}\right] \ .
\nonumber\\
&& {F^-(s_b,t)\over s_b-u_b} ={8\pi m_+ a_0^-\over\mpd-\mmd}
+\upi\intpik ds' \,\Bigg\{ 
{1\over \xlssb+s't}\im F^-(s',t'_b)-{1\over\xlss} \im F^-(s',0)\nonumber\\
&&\phantom{{F^-(s_b,t)\over s_b-u_b} =h^-(b)+\upi\intpik ds' }
-{1\over\xlssb} \im [F^-(s',t'_b)-F^-(s',0)] \Bigg\}\nonumber\\
&&\phantom{{F^-(s_b,t)\over s_b-u_b} =h^-(b)}
+{t\over2\pi}\intpipi {dt'\over t'(t'-t)} \im {G^1(t',s'_b)\over s'_b-u'_b}
\ena
These representations will allow us to perform projections on the 
$t$-channel partial waves for $t\ge 4\mpid$.

\subsection{RS equations for $f_l^I(s)$}

RS equations can now be obtained by performing the partial-wave
projections of the dispersive representations obtained above. Projecting
eqs.~\rf{fixtfin} on the $l=0,1$ $\pi K\to\pi K$ amplitude we get the
first four equations,
\bea\lbl{royfirst}  
&&\re f_l^\udemi(s)= k^\udemi_l(s)
\nonumber\\ 
&&\qquad +\upi\pvint\mpd{\infty} ds'\,\sum_{l'=0,1}
\Bigg\{ \left(\delta_{ll'}{\xls\over(s'-s)\xlss}
-\utier K^\alpha_{ll'}(s,s')\right)
\im f_{l'}^\udemi(s')
+\qtier K^\alpha_{ll'}(s,s') \im f_{l'}^\tdemi(s') \Bigg\}
\nonumber\\ 
&&\qquad +\upi\int_{4\mpid}^{\infty} dt' \Big\{ K^0_{l0}(s,t') \im g_0^0(t')
+2 K^1_{l1}(s,t') \im g_1^1(t')\Big\}
+d^\udemi_l(s)
\nonumber\\
&&\re f_l^\tdemi(s)= k^\tdemi_l(s) 
\nonumber\\
&&\qquad +\upi\pvint\mpd{\infty} ds'\,\sum_{l'=0,1}
\Bigg\{ \left(\delta_{ll'}{\xls\over(s'-s)\xlss}
+\utier K^\alpha_{ll'}(s,s')\right)
\im f_{l'}^\tdemi(s')
+\dtier K^\alpha_{ll'}(s,s') \im f_{l'}^\udemi(s') \Bigg\}
\nonumber\\ 
&&\qquad +\upi\int_{4\mpid}^{\infty} dt' \Big\{ K^0_{l0}(s,t') \im g_0^0(t')
- K^1_{l1}(s,t') \im g_1^1(t')\Big\} + d^\tdemi_l(s)\ .
\ena

The domain of validity in $s$ of these equations is given by 
eq.~\rf{svalid} below. In these equations,  
the terms $k^I_l(s)$ contain the contributions associated with the
subtraction constants,
\bea
&&k_0^I(s)={1\over2}\mpl a_0^I+{\xls\over32\pi s}\left(-b^+ +(-3I+{7\over2})\,
{8\pi\mpl a_0^-\over \mpd-\mmd}\,{3s+\mmd\over s-\mmd}\right)
\nonumber\\
&&k_1^I(s)={\xls\over96\pi s} \left(b^+ + (-3I+{7\over2})\,
{8\pi\mpl a_0^-\over \mpd-\mmd}\right)\ .
\ena

The equations involve three kinds of kernels $K^\alpha_{l l'}(s,s')$,
$K^I_{l l'}(s,t')$, and
$K^\sigma_{l l'}(s,s')$ (which appear only in the driving terms
$d^I_l$). 
The kernels $K^\alpha_{l l'}$ read, for $l,\ l'=0,\ 1$,
\bea
&&K^\alpha_{00}(s,s')=-{\xls+2s(s'-s)\over2s\xlss}+ L(s,s')
\nonumber\\
&&K^\alpha_{01}(s,s')={3(\xls+2s(s'+s))\over2s\xlss}-{3(\xlss+2ss'-2\Deld)
\over\xlss} L(s,s')
\nonumber\\
&&K^\alpha_{10}(s,s')={\xls^2+12s^2\xlss\over6s\xls\xlss}-{(\xls+2ss'-2\Deld)
\over\xls} L(s,s')
\nonumber\\
&&K^\alpha_{11}(s,s')=-{12s^2(\xlss+2ss'-2\Deld)+\xls^2\over2s\xls\xlss}
\nonumber\\
&&\phantom{K^\alpha_{11}(s,s')=-12s^2 }
+{3(\xls+2ss'-2\Deld)(\xlss+2ss'-2\Deld)\over\xls\xlss} L(s,s')
\ena
with
\be
L(s,s')={s\over\xls}\left[\log(s'+s-2\Sig)-
                          \log\left(s'-{\Deld\over s}\right)\right]
\en
Next, the kernels $K^0_{0l'}$, $K^0_{1l'}$ (with $l'$ even) read,
\bea
&&K^0_{0 l'}(s,t')={2l'+1\over\sqrt3}(q'_\pi q'_K)^{l'}{s\over\xls} \left\{
\log \left(1+{\xls\over st'}\right)-{\xls\over st'}
\left(1-{\xls\over 2 st'}\right)\right\}
\nonumber\\
&&K^0_{1 l'}(s,t')={2l'+1\over\sqrt3}(q'_\pi q'_K)^{l'}{s\over\xls}\left\{
\left(1+{2st'\over\xls}\right)\log \left(1+{\xls\over st'}\right)
-2-{1\over6}\left( {\xls\over st'}\right)^2\right\}\ .
\ena
Finally, the kernels $K^1_{0l'}$, $K^1_{1l'}$ (with $l'$ odd) read 
\bea
&&K^1_{0 l'}(s,t')={\sqrt2(2l'+1)\over 8} (q'_\pi q'_K)^{l'-1}\left\{
{s(2s-2\Sig+t')\over\xls} 
\left[\log \left(1+{\xls\over st'}\right)-{\xls\over st'}\right]
+{\xls\over2s t'} \right\}
\nonumber\\
&&K^1_{1 l'}(s,t')={\sqrt2(2l'+1)\over 8} (q'_\pi q'_K)^{l'-1}\times
\nonumber\\
&&\phantom{K^1_{1 l'}(s,t')=}
\left\{ {s(2s-2\Sig+t')\over\xls} 
\left[\left(1+{2s t'\over\xls}\right)\log \left(1+{\xls\over st'}\right)
-2\right] - {\xls\over6s t'}  \right\} \ .
\ena
The analyticity properties of the partial-wave amplitudes $f_l^I(s)$
were established in ref.~\cite{macdowell}. They can be recovered by considering
the various kernels. In particular, the circular cut is generated by
the kernels $K^I_{l l'}(s,t')$.

The terms $d^I_l(s)$ are the so-called driving terms in which 
the contributions from the partial waves with $l'\ge 2$ are collected
\bea
&&d^I_l(s)=
\ \upi\int_\mpd^{\infty} ds'\,\sum_{l'\ge 2}
\Bigg\{ \left(K^\sigma_{ll'}(s,s')+\dtier (I-1)\,K^\alpha_{ll'}(s,s')\right)
\im f_{l'}^\udemi(s')
\nonumber\\
&&\phantom{d^I_l(s)=
\ \upi\int_\mpd^{\infty} ds'\,\sum_{l'\ge 2}
\Bigg\{ \left(K^\sigma_{ll'}(s,s')\right) } 
+\utier (-2I+5)\, K^\alpha_{ll'}(s,s')\im f_{l'}^\tdemi(s')\Bigg\}
\nonumber\\ 
&&\ +\upi\int_{4\mpid}^{\infty} dt' \sum_{l'\ge 1}
\Big\{ K^0_{l 2l'}(s,t') \im g_{2l'}^0(t')
+(-3I+{7\over2}) K^1_{l 2l'+1}(s,t') \im g_{2l'+1}^1(t')\Big\} \ .
\ena
The kernels $K^\sigma_{l l'}(s,s')$ appear in the driving terms only;
the first few which are non-vanishing read
\bea
&&K^\sigma_{02}(s,s')={5 \xls\over s (\xlss )^2 }
\nonumber\\
&&K^\sigma_{03}(s,s')={-35 (\xls )^2 s'(s s'-\Delta^2)\over 3 s^2 (\xlss)^3}
\nonumber\\
&&K^\sigma_{13}(s,s')={7 \xls (s s'-\Delta^2)( (s+s')(s s'+\Delta^2)
-4 s s' \Sig)\over 3 s^2 (\xlss)^3 }\ .
\ena

\subsection{RS equations for $g_0^0(t)$, $g_1^1(t)$}

In order to obtain a closed system of equations we now need two equations
yielding the real parts of $g_0^0(t)$ and $g_1^1(t)$ valid
for positive values of $t$. They can be obtained 
from the family of fixed$-us$ DR's of eqs.~\rf{hypdrbfin}. Using the relation
between the cosine of the 
$t$-channel scattering angle $z_t$ and the parameter $b$,
\be
z^2_t={(2\Sig-t)^2-4b\over (2\Sig-t)^2-4\Deld}\ ,
\en
the projection is  carried out  by using
\bea\lbl{gproject}
&&g_0^0(t)={\sqrt3\over16\pi}\int_0^1 dz_t\, F^+(s_b,t)\nonumber\\
&&g_1^1(t)={4\sqrt2\over16\pi} \int_0^1 dz_t\, z^2_t\, 
{F^-(s_b,t)\over s_b-u_b}\ .
\ena
This yields the following two equations for $g_0^0$, $g_1^1$,
\bea\lbl{roylast}
&& g_0^0(t)={\sqrt3\mpl\over2}\left( a_0^+ +{t a_0^-\over \mpd-\mmd}\right)
+{t\over\pi}\intpipi {dt'\over t'}{\im g_0^0(t')\over t'-t}\nonumber\\
&&\phantom{g_0^0(t)=}
-{3\sqrt6\over8}{t\over\pi}\intpipi {dt'\over t'} \im g_1^1(t')\nonumber\\
&&\phantom{g_0^0(t)=}
+\sum^1_{l'=0} \upi\intpik ds' \left[ G^+_{0 l'}(t,s') \im  f^+_{l'}(s')
+t\, G^-_{0 l'}(t,s') \im  f^-_{l'}(s')\right]  + d_0^0(t)\ .
\nonumber\\
&&g_1^1(t)={2\sqrt2 \mpl a_0^-\over3(\mpd-\mmd)} +{t\over\pi}\intpipi
{dt'\over t'}{\im g_1^1(t')\over (t'-t)}\nonumber\\
&&\phantom{g_1^1(t)=}
+\upi\intpik ds' \,[ G^-_{1 0}(t,s') \im f^-_0(s') 
+G^-_{1 1}(t,s') \im f^-_1(s')]
+d_1^1(t) \ .
\ena
The two equations \rf{roylast} together with the four equations \rf{royfirst}
form a complete set of Roy-Steiner type equations. The domain of validity
of the equations for $g_0^0$, $g_1^1$ is given in eq.~\rf{tvalid} below.

The equation for $g_0^0$ involves three kinds of kernels: $G^\pm_{0 l'}(t,s')$,
$G^I_{0 l'}(t,t')$. The kernels $G^\pm_{0 l'}(t,s')$ have the following
form
\bea
&&G^+_{0 l'}(t,s')=\sqrt3(2l'+1)\left\{
{2G(x)\over s'-\Sigma+t/2} P_{l'}(z_{s'})
-{(2s'-2\Sigma+t)\over\xlss} A_{l'}(t,s') -{t  B_{l'}(t,s')\over\xlss}\right\}
\nonumber\\ 
&& G^-_{0 l'}(t,s')= \sqrt3(2l'+1){B_{l'}(t,s')\over\xlss}
\ena
where $P_{l'}$ are Legendre polynomials and
\be\lbl{Gx}
G(x)={{\rm arctanh}(x)\over x}, \quad 
x={\sqrt{Q_t}\over 2s'-2\Sig+t },\quad  
Q_t=(t-4\mpid)(t-4\mkd)\ .
\en
We collect below the expressions for the first few of the 
terms $A_l(t,s')$, $B_l(t,s')$
\be
\begin{array}{ll}
A_0(t,s')= 1,\ &B_0(t,s')=-1 \\
A_1(t,s')= 1,\ &B_1(t,s')= 1 \\
A_2(t,s')= 1+{\displaystyle 6s't\over\displaystyle\xlss},\ 
&B_2(t,s')= -\left(1+{\displaystyle Q_t\over\displaystyle \xlss}\right)\\
A_3(t,s')=1+{\displaystyle 10s't\over\displaystyle\xlss}
+{\displaystyle 10s't(-Q_t+6s't)\over\displaystyle 3\lambda^2_{s'}}
\ , &B_3(t,s')=1+{\displaystyle 5Q_t\over \displaystyle3\xlss}
+{\displaystyle 2Q^2_t\over\displaystyle 3\lambda^2_{s'}} \ .
\end{array}
\en
Lastly, we quote a few of the kernels $G^0_{0 2l'}(t,t')$,
$G^1_{0 2l'+1}(t,t')$,
\bea
&& G^0_{0 2}(t,t')={5\over16}(t'+t-4\Sig)\nonumber\\
&& G^0_{0 4}(t,t')={3\over256}(t'+t-4\Sig)\>
[3t'(t'-4\Sig)-7t(t-4\Sig)-64\mpid\mkd]
\nonumber\\
&& G^1_{0 3}(t,t')={-7\sqrt6\over384}\>[3t'(t'-4\Sig)-5t(t-4\Sig)
-32\mpid\mkd]\ .
\ena
In the RS equation for $g_1^1$, eq.~\rf{roylast}, one finds two
kinds of kernels $G^-_{1 l'}(t,s')$ and $G^1_{1 l'}(t,t')$. 
The kernels $G^-_{1 l'}(t,s')$ have a structure similar to 
$G^\pm_{0 l'}(t,s')$ encountered above,
\be
G^-_{1 l'}(t,s')=4\sqrt{2}(2l'+1)\left\{ {F(x)\over (s'-\Sig+t/2)^2}
-{1\over3\xlss}+C_{l'}(t,s')\right\}
\en
with
\be\lbl{Fx}
F(x)={1\over x^2} ( G(x)-1 )\ .
\en
$G(x)$ is defined in eq.~\rf{Gx} and both $F$ and $G$ are
smooth functions around 0. The pieces $C_{l'}(t,s')$ 
vanish for $l'=0,1$ and, for $l'=2,3$, read
\be
C_2(t,s')= -{2s't\over\xlss^2},\quad
C_3(t,s')=
{2s't\,[ (2s'-2\Sig+t)^2-9\xlss-14s't]\over3\xlss^3} \ .
\en
Finally, we display the first few  kernels $G^1_{1 2l'+1}(t,t')$ 
\bea
&&G^1_{1 3}(t,t')={7\over48}(t'+t-4\Sig)\nonumber\\
&&G^1_{1 5}(t,t')={11\over3840}(t'+t-4\Sig)
\big[ 5t'(t'-4\Sig )-9t(t-4\Sig) -64\mpid\mkd\big]\ .
\ena
These  kernels are seen to be polynomials in $t$, $t'$. 

The driving terms, $d_0^0(t)$, $d_1^1(t)$, in eqs.~\rf{roylast}  have the
following expressions
\bea
&&d_0^0(t)=\sum^\infty _{l'=2} \Big\{
\upi\intpik ds' \left[ 
     G^+_{0 l'}(t,s') \im f^+_{l'}(s')
+t\> G^-_{0 l'}(t,s') \im f^-_{l'}(s')\right] \nonumber\\
&&\phantom{d_0^0(t)= }
+{t\over\pi}\intpipi {dt'\over t'} \left[ 
 G^0_{0 2l'-2}(t,t')\im  g^0_{2l'-2}(t')
+G^1_{0 2l'-1}(t,t')\im g^1_{2l'-1}(t')\right]\,\Big\} 
\nonumber\\
&&d_1^1(t)=\sum^\infty_{l'=2}\Big\{ \upi\intpik ds' G^-_{1 l'}(t,s') 
\im f^-_{l'}(s') + {t\over \pi}\intpipi {dt'\over t'}\, 
G^1_{1 2l'-1}(t,t') \im g^1_{2l'-1}(t')
\Big\}\ .
\ena
This completes the derivation of a system of equations of the Roy-Steiner
type for $\pi K$ scattering. Let us now discuss the domain of validity
of these equations.

\section{Domains of validity} \label{sec:validity}

It is important to assess precisely the domains of validity of the
dispersive representations discussed in the preceding section. For this 
purpose, we will adapt the  methods reviewed by H\"ohler for 
the $\pi N$ system~\cite{hohler}. 
The discussion is based on the assumption that the scattering amplitudes
satisfy the Mandelstam double spectral representation~\cite{mandelstam0}, 
i.e., a spectral 
representation in terms of two variables which involves 
three spectral functions
$\rho_{st}(s',t')$, $\rho_{ut}(u',t')$ and $\rho_{us}(u',s')$. The boundaries
of the support of these spectral functions are shown in
fig.\fig{mandelstam}. 
This representation and the expressions for these boundaries are
obtained from the consideration of box diagrams (see for
instance~\cite{iz}). 
For the $\pi K$ amplitude, the ${st}$ boundary is described 
by the two equations
\bea\lbl{Bst}
&&(t-4\mpid)\lambda(s,\mkd,4\mpid)-16m^4_\pi(s+3\mkd-3\mpid)=0
\nonumber\\
&&(t-16\mpid)\lambda(s,\mkd,\mpid)-64m^4_\pi s=0
\ena
(the $ut$ boundary is obtained by replacing $s$
by $u$) while the ${us}$ boundary is defined by the following set
of equations
\bea\lbl{Bus}
&&\lambda(u,\mkd,4\mpid)\lambda(s,\mkd,\mpid)-16\mpid\mkd us
+16\mpid\Delta^2 (\mkd-t)=0
\nonumber\\
&&\lambda(u,\mkd,\mpid)\lambda(s,\mkd,4\mpid)-16\mpid\mkd us
+16\mpid\Delta^2 (\mkd-t)=0\ ,
\ena
with
\be
\lambda(x,y,z)=   x^2+y^2+z^2-2xy-2xz-2yz\ .
\en 

\begin{figure}[abt]
\epsfysize=9cm
\begin{center}
\psfrag{boundaries}{$Mandelstam\ boundaries$}
\psfrag{rhost}{$\rho_{st}$}
\psfrag{rhout}{$\rho_{ut}$}
\psfrag{rhous}{$\rho_{us}$}
\psfrag{smpid}{$s\ [m^2_\pi]$}
\psfrag{umpid}{$u\ [m^2_\pi]$}
\psfrag{tfour}{$t = 4$}
\psfrag{tsixteen}{$t = 16$}
\psfrag{tfourtyeight}{$t = -48$}
\includegraphics{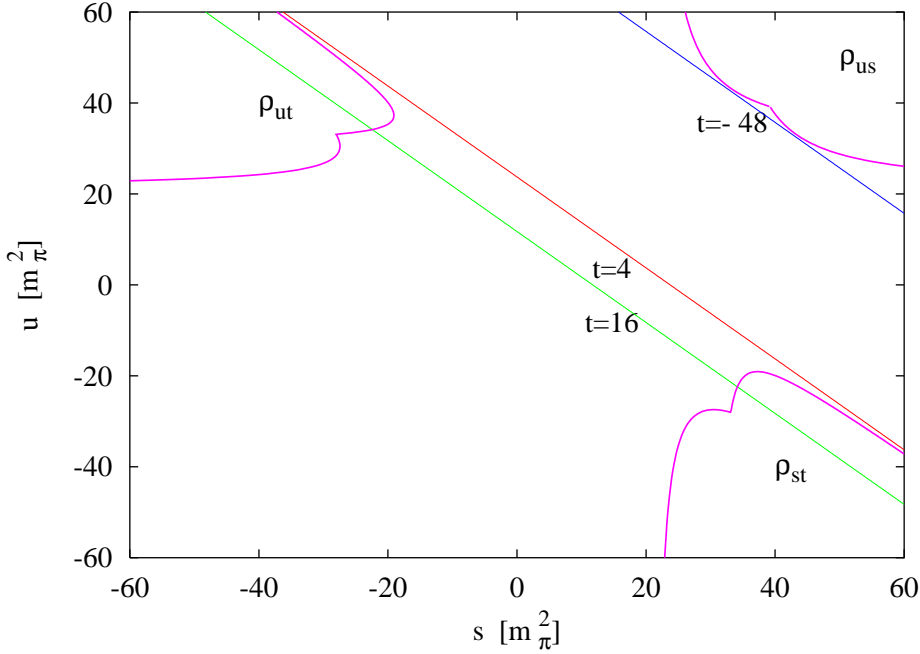}
\caption{ \sl Boundaries of the support of the Mandelstam double spectral
functions for the $\pi K$ system. The variables $s$, $t$, $u$ are 
displayed in units of $m^2_\pi$.}
\lblfig{mandelstam}
\end{center}
\end{figure}

Let us consider first the fixed$-t$ DR's. The spectral
functions arising in these DR's must be real, which 
implies that the lines of constant $t$ must not cross
the double-spectral boundaries. From fig.~\fig{mandelstam} 
one sees that this condition confines $t$ in the region,
\be
-48\mpid < t < 4\mpid\ ,
\en
where the lower bound comes from the boundary associated with $\rho_{us}$ 
and the upper bound from the one associated with $\rho_{st}$. 
The second restriction on the domain of validity arises
from the fact that the spectral function $\im F(s',t)$ is needed in an
unphysical region (except if $t=0$) and must thus
be defined using the partial-wave expansion. The domain of convergence
of this expansion is the large Lehman ellipse (see for instance~\cite{iz}). 
In terms of the cosine of the $s$-channel
scattering angle $z_{s'}$, this ellipse has focal points $z_{s'}=\pm1$
and it is limited by the $st$ spectral boundary,
\be
z^{\rm max}_{s'}=1+{2 s' T_{st}(s')\over \lambda_{s'} }\ .
\en 
The function $T_{st}(s)$ is obtained by solving eq.~\rf{Bst} 
which describes the $st boundary $for $t$ as a function of $s$. 
The point $-z^{\rm max}_{s'}$ of the 
ellipse corresponds to another value of $t$ given by 
$T'_{st}(s)= -{\lambda_s/ s} - T_{st}(s)$. For each value of $s'$, 
the convergence of the partial-wave expansion is ensured if 
$-z^{\rm max}_{s'}\leq z_{s'}\leq z^{\rm max}_{s'}$, i.e.,
$T'_{st}(s')<t<T_{st}(s')$.
{ The $us$ boundary provides another similar constraint, but it
turns out to be weaker than that obtained from the $st$ boundary.}
The conjunction of the two constraints (reality of the spectral
functions and convergence of the partial expansion) 
leads to the fact that the fixed-$t$ dispersion relation for $\pi K$ scattering
is valid in the range
\be\lbl{trange}
-23\mpid < t < 4\mpid\ .
\en

A similar discussion can be carried out for the set of dispersion relations
with $us$ fixed, $us=b$. Firstly,
the criterion that the hyperbolas $us=b$ do not intersect a spectral
function boundary yields
\be\lbl{brange0}
-700 m^4_\pi < b < 1420 m^4_\pi
\en
where the lower bound comes from the $st$ boundary and the upper bound
from the $us$ boundary. For the hyperbolic DR's, the spectral functions
$\im F^\pm (s',t'_b)$, $\im G^I(t',s'_b)$ are also needed in unphysical regions
(unless $b=\Delta^2$), so that the values of $b$ must be restricted to
ensure the convergence of the partial-wave expansion. Considering the Lehman
ellipse related to $\im F^\pm (s',t'_b)$ restricts the range to
\be\lbl{brange}
-700 m^4_\pi < b < 450  m^4_\pi\ ,
\en
and no further restriction arises from the Lehman ellipse related to 
$\im G^I(t',s'_b)$.

We can now derive the ranges of validity of the RS equations, which
are obtained by projecting the DR's over partial waves. 
Let us start with the fixed-$t$ DR's, 
the projection over $\pi K$ partial waves is legitimate provided
the range of integration of eq.~\rf{projects}
is included inside the range of validity in $t$ of the DR's. 
One deduces that the RS equations for
$s$-channel partial waves~\rf{royfirst} are valid for
\be\lbl{svalid}
3 \mpid \le s \le 48 \mpid\ .
\en
In the same way, the projection on $\pi\pi\to K\bar{K}$ partial waves
is allowed only if the range of integration of eq.~\rf{gproject} lies
within the range of validity in $b$ of the fixed$-us$ DR's. The last
two RS equations eq.~\rf{roylast} are thus valid for:
\be\lbl{tvalid}
-15 \mpid \le t \le 70 \mpid \ .
\en
The range of validity in $t$ is significantly larger than that in $s$.
This difference stems from Bose symmetry, which applies only to the
$\pi\pi\to K\bar{K}$ channels and implies that only
even (odd) partial waves appear when the isospin is zero (one).
Thus, the $t$-channel projections can be obtained 
by integrating over the limited range $0\le z_t\le 1$, whereas 
the projection on $s$-channel partial waves requires integrating 
over the whole range $-1\le z_s\le 1$.
One notes that it is possible to project the hyperbolic DR's over
$s$-channel partial waves as well. However, the resulting partial-wave 
equations are valid in the range $s\le 43 \mpid$, 
which is somewhat smaller than the range of validity of the partial-wave
equations obtained  from the fixed$-t$ DR's.

\section{Experimental input}

In the previous sections, we have derived a set of RS equations for the
$s$-channel partial waves for $I={1\over2},{3\over2}$ and $l=0,1$,
and the $t$-channel partial waves for $(I,l)=(0,0)$ and $(1,1)$, which
we call ``lowest'' partial waves from now on. 
Let us consider these equations in the ranges $\mpd\le s\le s_m$
and $4\mpid\le t\le t_m$. The upper limits of which $s_m$, $t_m$ 
(which will be taken such that the equations are valid i.e. 
$s_m\le 48 m^2_\pi$, $t_m\le 70 m^2_\pi$)
will be called {\sl matching points}. A simple examination of the RS equations
shows that in order to be able to solve for the lowest partial waves below
the matching points the following input must be provided: 1) the imaginary
part of the lowest partial waves for $s\ge s_m$, $t\ge t_m$, 2) the imaginary
parts of the $l\ge 2$ partial waves above the thresholds and 3) the phases
of $g_0^0(t)$, $g_1^1(t)$ in the range $4\mpid\le t\le t_m$. We will discuss
below the experimental status of this input.

For the $s$-channel partial waves, we choose the matching point at the
border of the range of validity:
\begin{equation}
s_m=0.935\  {\rm GeV}^2\ .
\end{equation}
The reason for this choice is that the experimental data available
at present comes from production experiments. One expects the precision
to decrease as the energy goes down below 1 GeV. We will see, for instance,
that the $I={3\over2}$ $S$-wave phase shifts seem rather unreliable below
1 GeV. In the $t-$channel the range of validity extends, as we have seen,
up to $t_{val}\simeq 1.36$ GeV$^2$ and one could, in principle, choose the
matching point anywhere between the $K\Kbar$ threshold and $t_{val}$.
In practice, we choose a value slightly above the $K\bar{K}$ threshold
(see sec.~\ref{sec:numerical})
\begin{equation}
t_m=1.04\  {\rm GeV}^2
\end{equation}

For the lowest partial waves above the matching point, and for
the higher partial waves, we exploit experimental data at intermediate
energies 
\be
E\le \sqrt{s_2}=2.5\ {\rm GeV}
\en
and Regge models for $E> \sqrt{s_2}$. We aim
at determining the lowest partial waves below the matching point.
For this purpose, an additional information is needed concerning unitarity.
We will make the usual assumption that elastic unitarity holds 
exactly below the matching points~\cite{acgl}. 
In other terms, in the $\pi K$ channel
the possible couplings to $\pi\pi K$ and $\pi\pi\pi K$ are assumed to be
negligibly small in the low-energy region. For the $S$-wave 
the validity of elastic unitarity was observed experimentally 
up to the $\eta' K$ threshold. In principle,
the $P$-wave can couple to the $\pi\pi K$ state but no such coupling
has been detected for the $K^*$~\cite{pdg}, 
and potentially important two-body channels
like $K^*\pi$, $K\rho$ lie above the matching point. Similarly, in the 
$\pi\pi$ channel we assume that the coupling to $4\pi$ can be neglected 
below the $K\Kbar$ threshold.

We discuss now the experimental input used to solve the RS equations,
before explaining in detail their resolution.

\begin{figure}[th]
\begin{center}
\includegraphics[width=12cm]{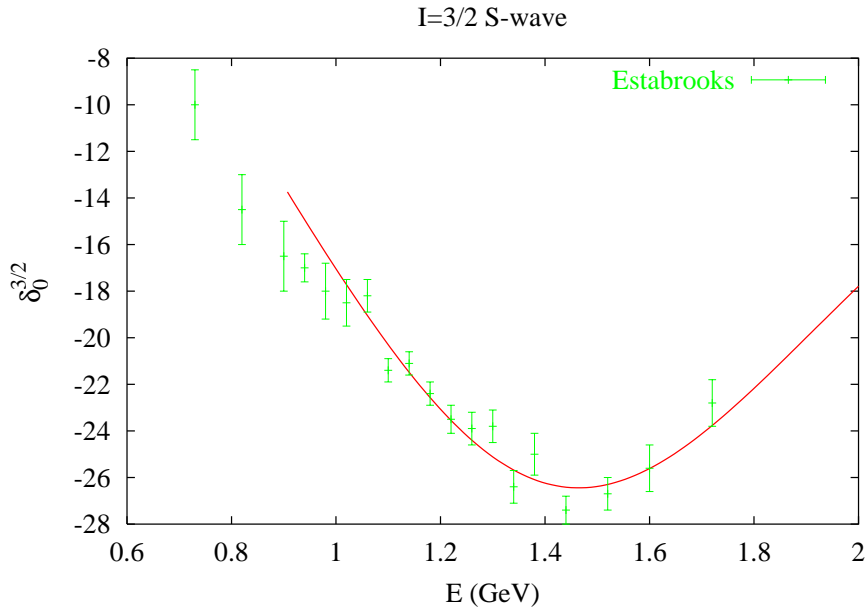}
\caption{\sl Experimental data from ref.~\cite{aston} for the $I={3\over2}$
$S$-wave phase shift  }
\lblfig{S32-wave}
\end{center}
\end{figure}

\subsection{$\pi K\to \pi K$ data }

Phase shift analyses of the $\pi K\to \pi K$ amplitude have been performed 
based on high-statistics production experiments $KN\to K\pi N$ by 
Estabrooks {\it et al.}~\cite{estabrooks} and by 
Aston {\it et al.}~\cite{aston}. Earlier results are much less
precise and  we will not use them in our analysis. 
The amplitude 
$\pi^+ K^+\to \pi^+ K^+$ which is purely $I={3\over2}$ has been 
measured by Estabrooks {\it et al.}~\cite{estabrooks}. 
In practice the $I={3\over2}$ phase shifts are
very small in the range $E\lapprox2$ GeV except for the $S$-wave. 
This phase shift is shown in fig.\fig{S32-wave}  together 
with our fit, where a simple parametrization with three parameters is used
\be
\tan\left(\delta_0^{3/2}(s)\right)={\alpha q\over 1+\beta q^2 +\gamma q^4}
\en
This parametrization is analogous to the one used in ref~\cite{jop}. 
Inelasticity is neglected in this channel. 

The amplitude 
$\pi^+ K^-\to \pi^+ K^-$ which involves the following isospin combination
\be
F^c\equiv F^{1/2}+{1\over2} F^{3/2}\ ,
\en 
was measured both in ref.~\cite{estabrooks} and
ref.~\cite{aston} -- the latter experiment has better statistics and covers
a larger energy range.
The amplitude $F^c$ can be expanded over partial-waves in the same way as
eq.~\rf{Fpw} and refs.~\cite{estabrooks,aston} provide the phase $\Phi_l(s)$
and the modulus $a_l(s)$ of these partial waves,
\be
f^c_l(s)\equiv \sqrt{2l+1}\, a_l(s) {\rm e}^{i\Phi_l(s)}\ .
\en
\begin{figure}[th]
\begin{center}
\includegraphics[width=12cm]{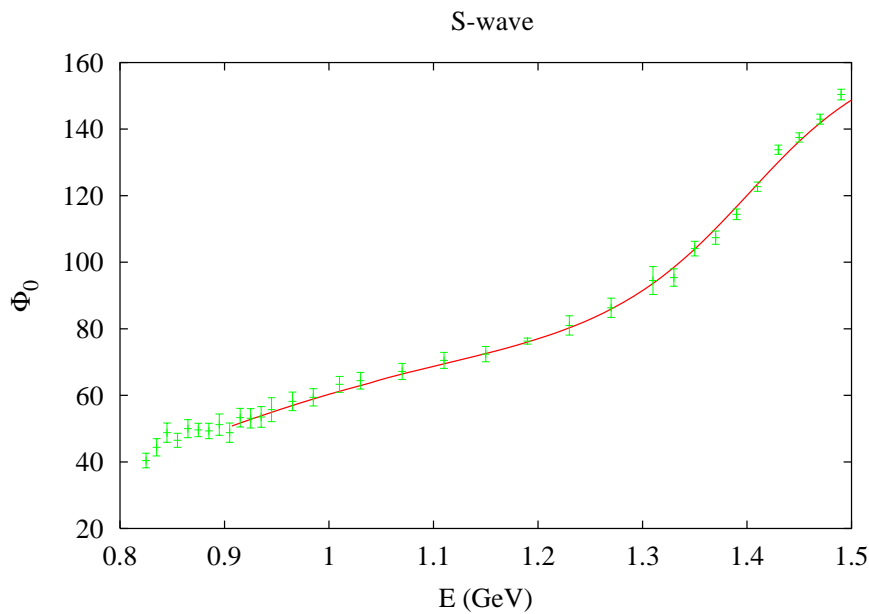}
\caption{\sl Experimental data from ref.~\cite{aston} for the phase $\Phi_0$ 
and the fit used in the calculations.}
\lblfig{S-wave0}
\end{center}
\end{figure}
\begin{figure}[thb]
\begin{center}
\includegraphics[width=12cm]{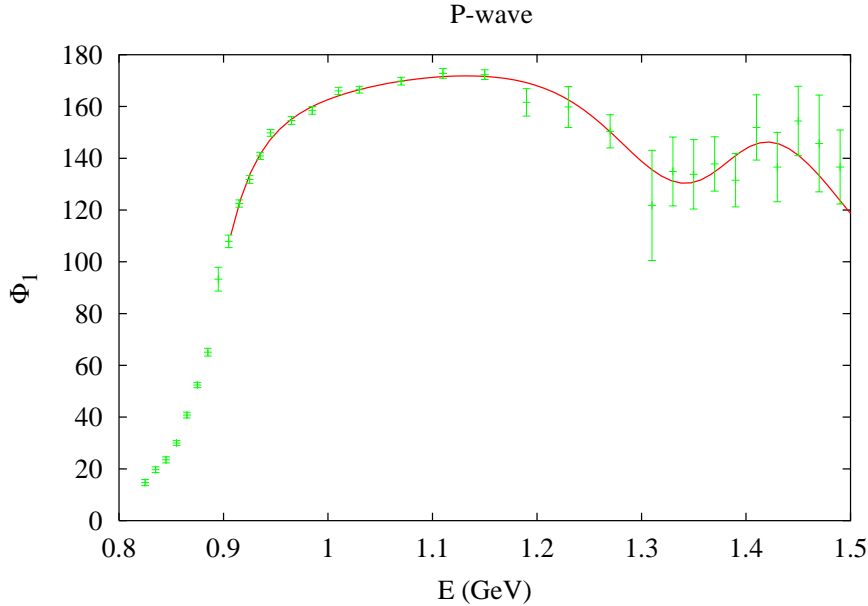}
\caption{\sl Experimental data from ref.~\cite{aston} for the phase $\Phi_1$ 
and the fit used in the calculations.}
\lblfig{P-wave0}
\end{center}
\end{figure}

Performing a combined fit of the $I={3\over 2}$ partial waves~\cite{estabrooks}
and of the
parameters $a_l$, $\Phi_l$~\cite{estabrooks,aston} 
one can separate the two isospin partial waves.
The data of Aston {\it et al.} for the phases $\Phi_0$ and $\Phi_1$ 
and our fits  are displayed in figs.~\fig{S-wave0} and \fig{P-wave0}
respectively in the range 
$0.9\le E\le 1.5$ GeV (this energy region plays an important role
in our analysis). The fits shown here correspond to a
parametrization of the partial-wave $S$-matrices
as products of Breit-Wigner $S$-matrices, allowing
for inelasticity in the  $I={1\over 2}$ amplitude to set in 
at the $\eta K$ threshold. 
Inelasticity is found to remain quite small up to $E\simeq 1.5$ GeV. 
We also tried different fits based on K-matrix parametrizations. 

The data of Aston {\it et al.} and the fits for both $a_l$ and $\Phi_l$ for 
$l=0$ to $l=5$ and energy up to $E=2.5$ GeV
are shown in figs.~\fig{SP-waves}, \fig{DF-waves} and  \fig{GH-waves}. 
At energies $E\ge 1.8$ GeV, ref.~\cite{aston} found two different
solutions A and B for the phase shifts, between which
we choose sol A (it was pointed out
in ref.~\cite{jop} that solution B violates the unitarity bound).
These fits allow us to compute the
relevant integrals up to $E=2.5$ GeV. Above that point, we use
a Regge-model parametrization discussed in sec.~\ref{sec:regge}.

\begin{figure}[htb]
\begin{center}
\includegraphics[width=15cm,height=6cm]{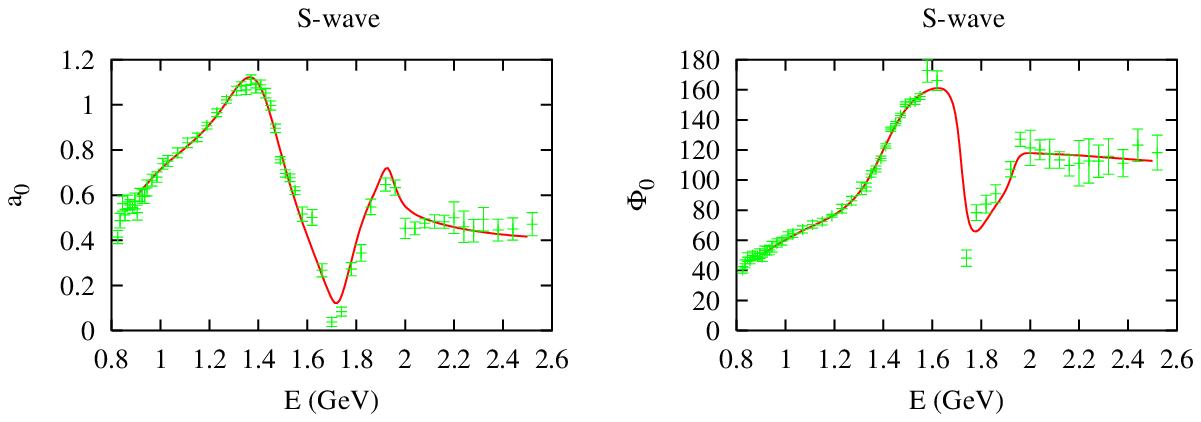}
\includegraphics[width=15cm,height=6cm]{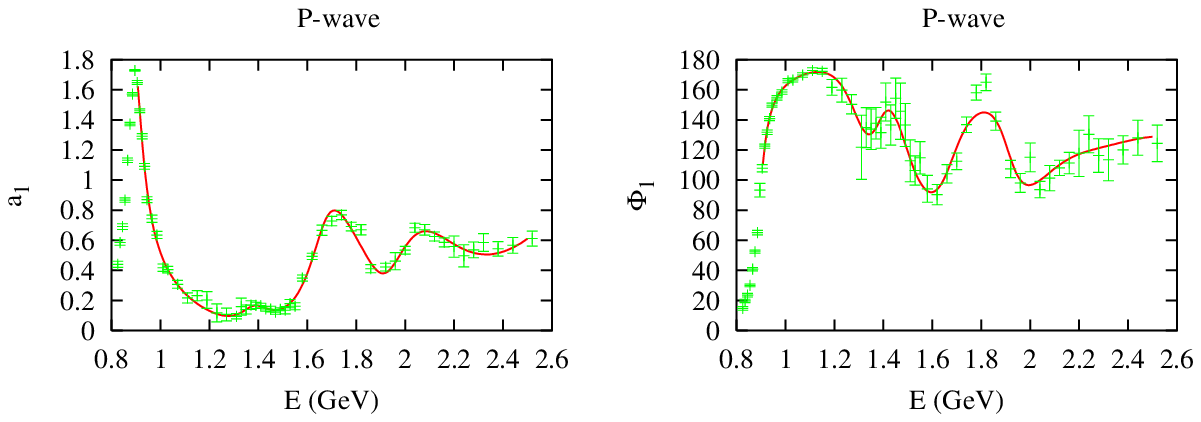}
\caption{\sl  Modulus and phase of the $S$- and  $P$-partial-waves 
amplitudes from ref.~\cite{aston}  and the fits 
in the region $0.9\le E\le 2.5$ GeV which are used
in the calculations.}
\lblfig{SP-waves}
\end{center}
\end{figure}

\begin{figure}[thb]
\begin{center}
\includegraphics[width=15cm,height=6cm]{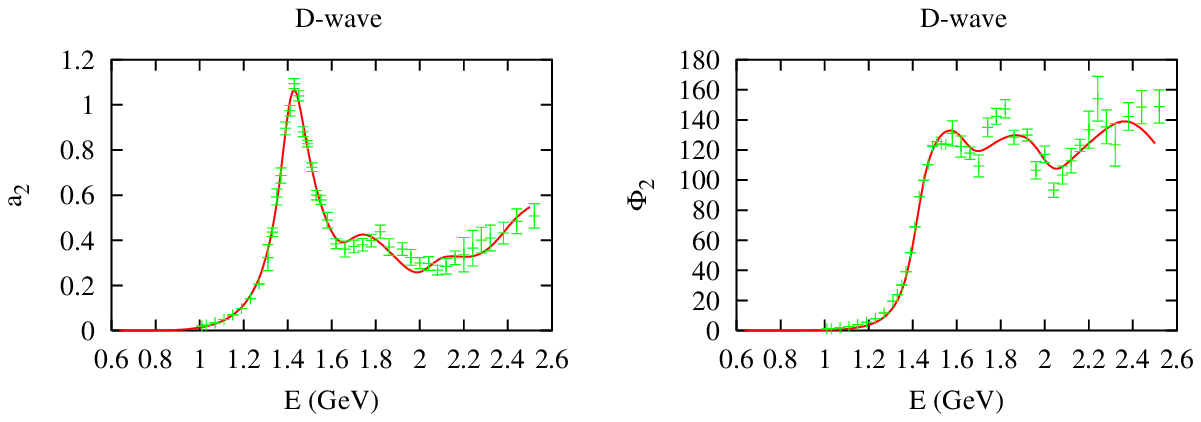}
\includegraphics[width=15cm,height=6cm]{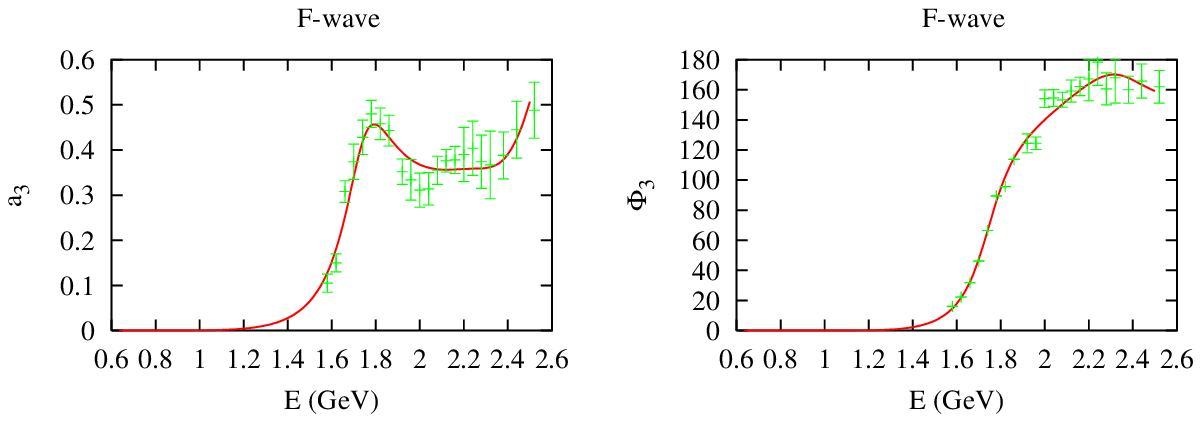}
\caption{\sl Same as fig.\fig{SP-waves} for the $D$- and $F$-partial waves.}
\lblfig{DF-waves}
\end{center}
\end{figure}
\begin{figure}[thb]
\begin{center}
\includegraphics[width=15cm,height=6cm]{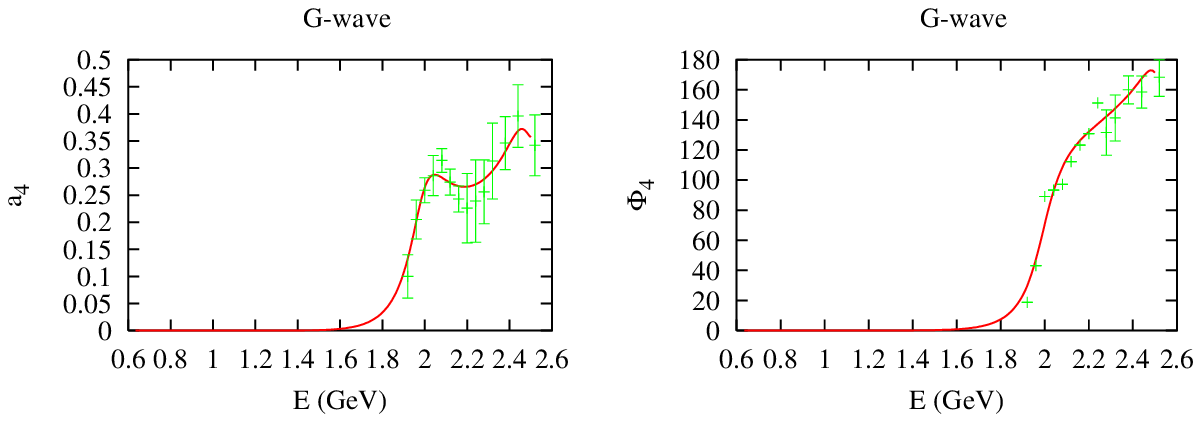}
\includegraphics[width=15cm,height=6cm]{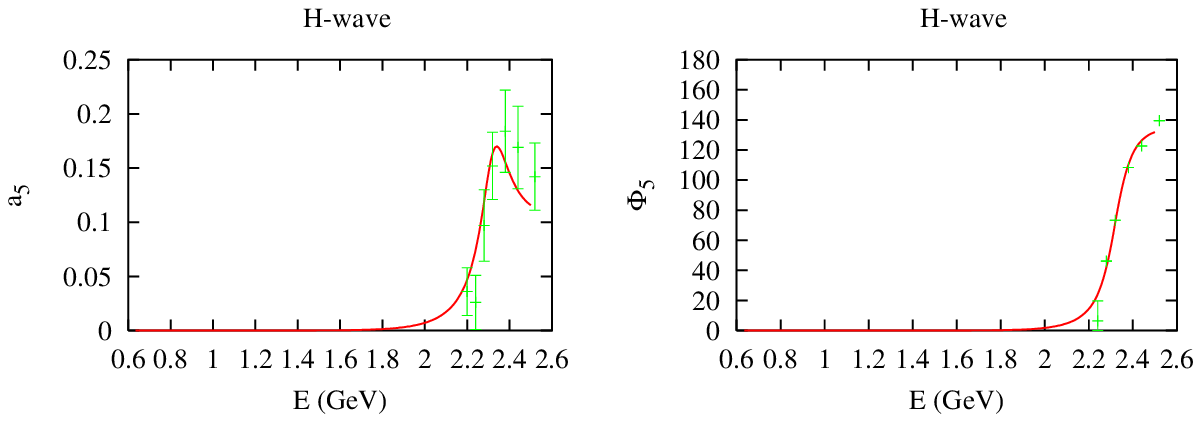}
\caption{\sl Same as fig.\fig{SP-waves}  for the $G$- and $H$-partial waves.}
\lblfig{GH-waves}
\end{center}
\end{figure}

\subsection{$\pi\pi\to K\overline{K}$ input }

For our purposes, a key role is played 
by the $l=0$ and $l=1$ $\pi\pi\to K\overline{K}$ amplitudes,
which can be determined from $\pi N\to K\Kbar N$
production experiments in the range $t\ge 4\mkd$.
We will make use of
the two high-statistics experiments described in
Cohen {\it et al.}~\cite{cohen} and
Etkin {\it et al.}~\cite{etkin,longacre}. 
The experiment of Cohen {\it et al.}~\cite{cohen} determines 
the charged amplitude
$\pi^+\pi^-\to K^+ K^-$, thereby providing results for both $g_0^0$ and 
$g_1^1$. There are several possible solutions but physical requirements
select a single one, called solution II b in ref.~\cite{cohen}. 
Close to the $K \Kbar$ threshold, the presence of the $l=1$ phase allows 
the authors to accurately determine the $l=0$ phase. 
The experiment of Etkin {\it et al.} concerns the amplitude 
$ \pi^+\pi^-\to K_S K_S$ which is purely $I=0$. 
Because of the absence of the $P$-wave in this channel, 
their determination of the
phase of $g_0^0$ close to the threshold (where the $D$-wave phase 
is very small)
is likely to be less reliable than that of ref.~\cite{cohen}. 
Their determination of the magnitude of $g_0^0$ close to the threshold 
disagrees with that of Cohen {\it et al.} and also with earlier 
experiments~\cite{oldpipikkbar}. 
Consequently, we make the choice to 
use the results of Etkin {\it et al.} only in the range $\sqrt{t}\ge 1.2$ GeV.

Our input for the phase of $g_0^0$ is determined as follows. 
Below the $K\Kbar$ threshold this phase is identical to the $\pi\pi$
phase shift because of the elastic unitarity assumption.
In the range $2m_\pi\le E\le 0.8$ GeV 
we use solutions of the $\pi\pi$ Roy
equations. Simple parametrizations were provided recently in 
refs.~\cite{acgl,dfgs}. We use the
parametrization of ref.~\cite{dfgs} together with the scattering lengths
corresponding to the ``extended'' fit,
with the central values $a_0^0=0.228$, $a_0^2=-0.0382$. 
In the range $E\ge 2m_K$ we perform
piecewise-polynomial fits of the data of refs.~\cite{cohen,etkin} and fixing
the threshold value to $\Phi_0^0=200\pm 15 $ degrees. This range is an
educated guess based on considering the data of Cohen {\it et al.} as well
as $\pi\pi$ data. Finally in the range $0.8 \textrm{ GeV}\le E\le 2m_K$ 
we perform a fit to the CERN-Munich data as given by 
Hyams {\it et al.}~\cite{hyams} and to
the polarized target production data recently analyzed 
by Kaminski {\it et al.}~\cite{kaminski}. For the modulus of $g_0^0$,
we have performed piecewise polynomial fits to the data of 
refs.~\cite{cohen,etkin}. The data and these fits are shown in 
fig.\fig{g00input}.

\begin{figure}[thb]
\begin{center}
\includegraphics[width=16cm]{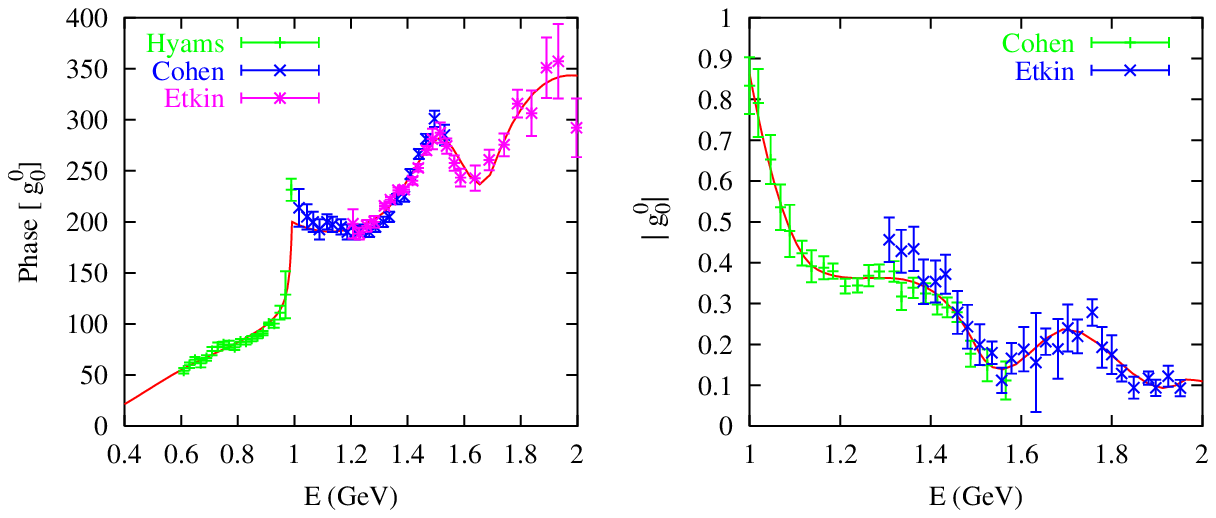}
\caption{\sl Inputs for the phase of $g_0^0$ above the $\pi\pi$ 
threshold and its modulus above the $K\Kbar$ threshold. The data are from
refs.~\cite{hyams,cohen,etkin}. }
\lblfig{g00input}
\end{center}
\end{figure}

As far as $g_1^1$ is concerned, we use the
experimental determination of the $\pi\pi$ $P$-wave phase 
in the range $2m_\pi \le E\le 2m_K$ obtained from the
pion vector form factor measured by CLEO~\cite{cleo}. 
This determination is compatible with
the results of the analysis of $\pi\pi$ Roy equations and has a 
comparable accuracy.
At larger energies, we use the experimental results from Cohen {\it et al.}
for the phase and the magnitude of $g_1^1$. The whole energy range 
where data are available can be fitted using the following form
\bea\lbl{cleo}
&&g_1^1(t)= {C\over (1+r_1 q^2_\pi(t) )^\undemi (1+r_1 q^2_K(t) )^\undemi}
\times\nonumber\\
&&\phantom{g_1^1(t)= {C } }
\left\{ BW(t,m_\rho) + (\beta+ \beta_1 q^2_K(t)) BW(t,m_{\rho'}) +
(\gamma+ \gamma_1 q^2_K(t)) BW(t,m_{\rho''})\right\} 
\ena 
with
\be
BW^{-1}(t,m_V)=
m^2_V-t-i\sqrt{t}\>\Gamma_V {2G_\pi(t)+G_K(t)\over 2G_\pi(m^2_V)}
,\ G_P(t)=\sqrt{t}\left({2q_P(t)\over\sqrt{t}}\right)^3\ .
\en
Below the $K\Kbar$ threshold, $q^2_K(t)$ vanishes and  expression~\rf{cleo} 
reduces to the 
K\"uhn and Santamaria~\cite{KS} form used in ref.~\cite{cleo}. 
We take the values of the parameters $\beta$, $\gamma$, $m_\rho$, 
$m_{\rho'}$, $m_{\rho''}$  determined by CLEO and we fit the 
parameters $C$, $r_1$,
$\beta_1$, $\gamma_1$  to the data above the $K\Kbar$ threshold. 
The data and the fits are shown in fig.~\fig{g11input}.

\begin{figure}[thb]
\begin{center}
\includegraphics[width=16cm]{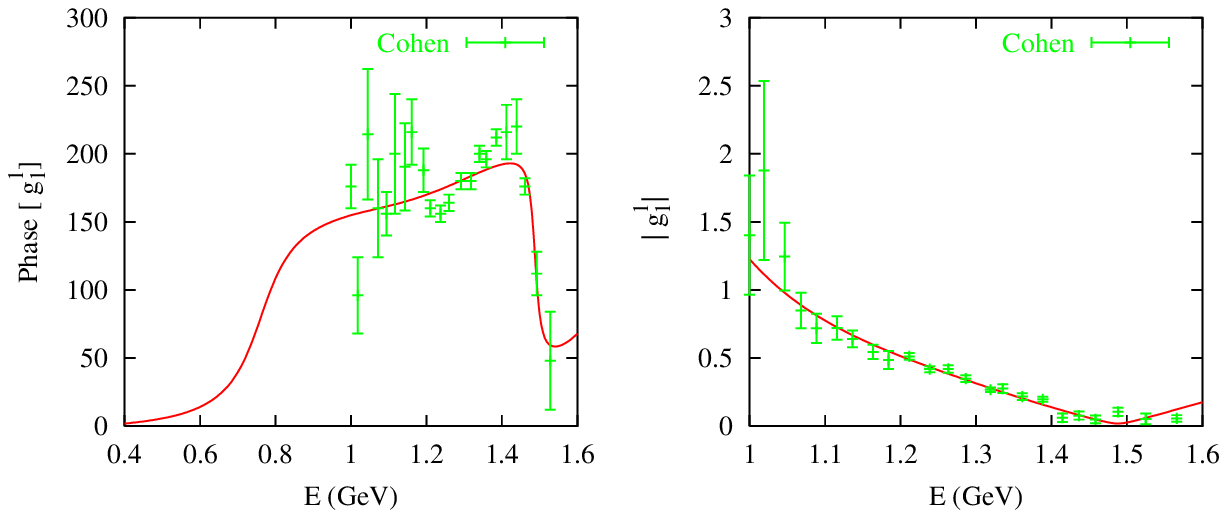}
\caption{\sl Input for the phase and the magnitude of $g_1^1$. The data
shown are from ref.~\cite{cohen} and the curve is the fit from eq.~\rf{cleo}.}
\lblfig{g11input}
\end{center}
\end{figure}

The amplitudes with $l\ge 2$ play a much less significant role
in our analysis and are suppressed at low energies.
They will be described
by simple Breit-Wigner parametrizations associated with the resonances
$f_2(1200)$, $f'_2(1525)$, $\rho_3(1690)$, $f_4(2050)$. 
Masses and partial decay
widths of these resonances were taken from the PDG~\cite{pdg}.

\subsection{Asymptotic regions} \label{sec:regge}

As discussed above, we can make use of the partial-wave expansion 
and experimental data up to energies $E=\sqrt{s_2}=2.5$ GeV for the 
$s$- as well as the  $t$-channel. Above that point we use a description of the
amplitudes based on Regge phenomenology. We will content
ourselves with very unsophisticated models because this 
energy region  turns out
to play a very minor role in our analysis.
In the regime $s'\to\infty$, $t$ fixed, we use the following expression
for the amplitudes suggested by dual 
models \`a la Veneziano~\cite{veneziano,lovelace,shapiro,kawarabayashi}
(where exact exchange degeneracy is built in)
\be\lbl{reggemin}
\im F^-(s',t)\vert_{\rm asy}\sim {\pi\lambda\over \Gamma(\ar +\alpha_1 t)}
(\alpha_1 s')^{\ar+\alpha_1 t} 
\en
and
\be
\im  F^+(s',t)\vert_{\rm asy}\sim \sigma s' {\exp}\left({bt\over2}\right)
+ \im  F^-(s',t)\vert_{\rm asy}\ .
\en
For the parameter $\lambda$ and the Pomeron parameters $\sigma$, $b$ 
we adopt values inspired by the discussion in App.~B.4
of ref.~\cite{acgl} with large errors 
\be
\lambda=14\pm 5\ ,\ 
\sigma= 5\pm 2.5 {\ \rm mb}\ ,\ 
b= 8\pm 3 \ {\rm GeV}^{-2}\ .
\en
The intercept and slope 
parameters of the Regge trajectories are determined from the 
experimental spectrum  of the $\rho$ and $K^*$ resonances,
\be
\ak=0.352,\quad \ar=0.475,\quad \alpha_1=0.882\ {\rm GeV^{-1}}\ .
\en
For illustration we compare in fig.~\fig{regge} the imaginary part of
$F^-(s,0)$ resulting from our fit to the experimental data and the Regge 
asymptotic form with  $\lambda=14$ .

Making use of this, it is easy to evaluate the contributions to the
various dispersive integrals in the range $[s_2,\infty]$.
In the fixed-$t$ DR's we obtain
\bea
&&F^+(s,t)\vert_{s_2}= {2(\xls+ s t)\over (s_2)^2}\left[
{\sigma s_2\over\pi} {\rm exp}{bt\over2}
+{\lambda\over (2-\ar-\alpha_1 t)\Gamma(\ar+\alpha_1 t)}
(\alpha_1 s_2) ^{\ar+\alpha_1 t}\right]\nonumber\\
&&\phantom{F^+(s,t)_{s_2}}
+{t^2\over (s_2)^2}\left[ {\lambda\over (2-\ak)\Gamma(\ak)}
(\alpha_1 s_2) ^\ak  \right]
\ena
and
\be\lbl{fminasy}
F^-(s,t)\vert_{s_2}= {s-u\over (s_2)^2}\left[ 
t\, {\lambda (\alpha_1 s_2)^\ak \over (2-\ak)\Gamma(\ak)} +
{\xls +s t\over s_2}{\lambda (\alpha_1 s_2)^{\ar+\alpha_1 t}
\over (3-\ar-\alpha_1 t)\Gamma(\ar+\alpha_1 t)}
\right]
\en
In the same manner we can obtain the asymptotic contributions in the 
amplitudes described through hyperbolic DR's
\bea
&&
{F^-(s_b,t)\over s_b-u_b}\Bigg\vert_{s_2}={\Deld-b\over (s_2)^3 }
{\lambda (\alpha_1 s_2)^\ar\over (3-\ar)\Gamma(\ar) }
+{t\over (s_2)^2 }
{\lambda (\alpha_1 s_2)^\ak\over \Gamma(\ak) }\Bigg[ {1\over2-\ak}
\nonumber\\
&&\phantom{{F^-(s_b,t)\over s_b-u_b}}
-{ \, b\, \alpha_1\over s_2(3-\ak)}
\left( \log(\alpha_1 s_2)-\psi(\ak)+{1\over3-\ak}\right) \Bigg]
\ena
and
\bea
&&F^+(s_b,t)\vert_{s_2}= {2(\Deld-b)\over (s_2)^2 }
\left[ {\sigma s_2\over\pi}+
{\lambda (\alpha_1 s_2)^\ar\over (2-\ar)\Gamma(\ar) }\right]
+{t(t-2\Sig)\over (s_2)^2 }
{\lambda (\alpha_1 s_2)^\ak\over \Gamma(\ak) }\Bigg[ {1\over2-\ak}
\nonumber\\
&&\phantom{{F^-(s_b,t)\over s_b-u_b}}
-{ \, b\, \alpha_1\over s_2(3-\ak)}
\left( \log(\alpha_1 s_2)-\psi(\ak)+{1\over3-\ak}\right) \Bigg]
\ena
To derive these contributions, we have used the following 
expression for $\im G^I(t',s'_b)$ in the regime where  $t'\to\infty$
\be
{1\over\sqrt6} \im G^0(t',s'_b)\vert_{asy}=
{1\over 2} \im G^1(t',s'_b)\vert_{asy}=
{\pi\lambda\over \Gamma(\ak ) }(\alpha_1 t')^{\ak } 
\left\{ 1 +{\alpha_1b\over t'}( -\log(\alpha_1 t')+\psi(\ak) )\right\}
\en
in which an expansion to first order in the parameter $b$ has been performed.

\begin{figure}[abt]
\begin{center}
\includegraphics[width=12cm]{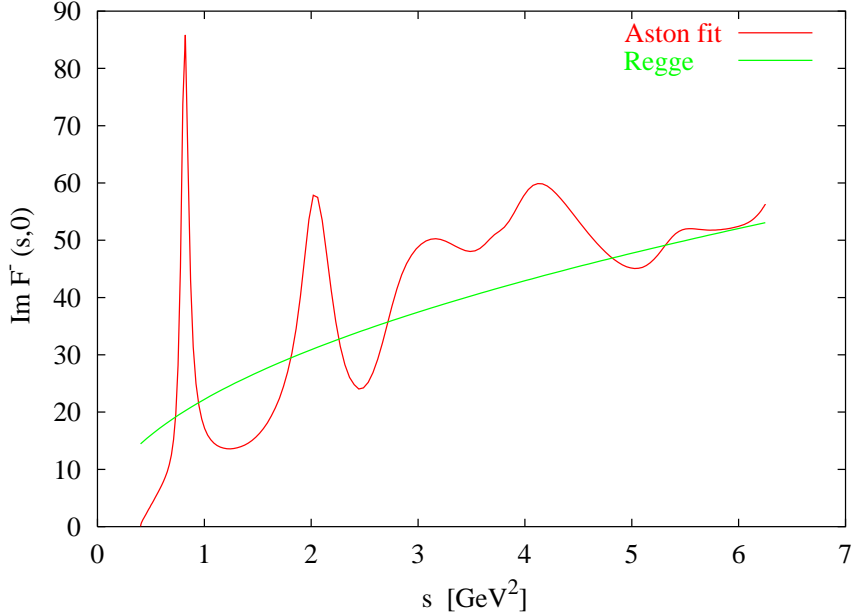}
\caption{\sl Comparison of $\im F^-(s,0)$ constructed from experimental 
data and the Regge asymptotic form eq.~\rf{reggemin}.}
\lblfig{regge}
\end{center}
\end{figure}

\section{Initial steps in the resolution}

\subsection{Solving for $g_0^0$, $g_1^1$}

We have now all the ingredients to solve the set of RS equations.
The first step consists in solving
eqs.~\rf{roylast} for $g_0^0$, $g_1^1$. This problem was discussed 
a long time ago~\cite{johannesson0,hedegaard} and
we recall the main ideas here
for completeness. Elastic unitarity implies
that the phases $\Phi^I_l$ of these amplitudes
\be
g^I_l(t)\equiv {\rm e}^{i\Phi^I_l(t)} \vert g^I_l(t)\vert
\en
can be identified with the $\pi\pi$ phase shifts $\delta^I_l$
in the unphysical region $t< 4m^2_K$ according to 
Watson's theorem~\cite{watson}, and therefore they are known in principle. 
In the physical region $t\ge 4m^2_K$ the phases are determined
from experiment as was discussed above. 

On the other hand, the modulus of the $t$-channel partial waves is not 
known below the $K\bar{K}$ threshold, and must be determined using 
the equations~\rf{roylast} satisfied by $g_0^0$ and $g_1^1$
which have the following simple form
\bea\lbl{roylast1}
&&g_0^0(t)=\Delta_0^0(t)+{t \over\pi}\int_{4m^2_\pi}^\infty 
{dt'\over t' }{\im  g_0^0(t')\over t'-t},
\nonumber\\ 
&&g_1^1(t)=\Delta_1^1(t)+{t\over\pi}\int_{4m^2_\pi}^\infty 
{dt'\over  t'}{\im  g_1^1(t')\over t'-t}\ .
\ena
In sec.~\ref{sec:validity}, we have shown that these relations can be used up 
to $t\simeq 1.4 \  {\rm GeV}^2$,
which includes the whole region inaccessible to experiment where
$\vert g_0^0\vert$, $\vert g_1^1\vert$ are needed.
The quantities $\Delta^I_l(t)$ are analytic functions  with a left-hand 
cut along the negative $t$ axis and no right-hand cut,
as can be easily verified using eqs.~\rf{roylast} 
and the explicit form of the kernels provided
in sec.~\ref{sec:ampldisp}. 
Therefore, determining the moduli $\vert g_l^I(t)\vert $ in the range 
$4\mpid\le t \le \tm$ from eqs.~\rf{roylast1} while the phase is known 
is a standard Muskhelishvili-Omn\`es 
problem~\cite{muskhelishvili,omnes}. The most general solution involves
arbitrary parameters, the number of which depend on the value of the 
phase at the matching point~\cite{muskhelishvili}. 
We have chosen $\tm$ to be slightly
larger than $4\mkd$. The $l=1$ phase $\Phi_1^1(\tm)$ is lower than
$\pi$, which implies that the solution for $g_1^1$ involves no 
free parameter. The $l=0$
phase, as we argued in the previous section, 
satisfies $\pi\le  \Phi_0^0( \tm)< 2\pi$,
such that one free parameter is involved in the solution. Let us recall
the explicit form of the solutions. One first introduces the Omn\`es
function
\be\lbl{omdef}
\Omega^I_l(t)=\exp\left({t\over\pi}\int_{4\mpid}^{\tm}
{\Phi^I_l(t')\,dt'\over t'(t'-t)}\right)\equiv 
{\Omega^I_l}_R(t)\exp [i \Phi^I_l(t)\theta(t-4\mpid)\theta(\tm-t)]
\en
where ${\Omega^I_l}_R(t)$ is real. Then, the solutions of eqs.~\rf{roylast1}
read
\bea\lbl{omformules}
&&g_0^0(t)= \Delta_0^0(t)+{t \Omega_0^0(t)\over \tm-t}\Bigg[\alpha_0 
+{t\over\pi}\int_{4\mpid}^{\tm} dt'\, {(\tm-t')\Delta_0^0(t')\sin\Phi_0^0(t')
\over {\Omega^0_0}_R(t')(t')^2 (t'-t)}
\nonumber \\
&&\phantom{\Delta_0^0(t)+{t \Omega_0^0(t)\over \tm-t}\Bigg[\alpha_0}
+{t\over\pi}\int_\tm^\infty dt'\,{(\tm-t')\vert g_0^0(t')\vert \sin\Phi_0^0(t')
\over {\Omega^0_0}_R(t')(t')^2 (t'-t)}\Bigg]
\ena
\bea\lbl{omformules1}
&&g_1^1(t)= \Delta_1^1(t)+t \Omega_1^1(t)\Bigg[ 
{1\over\pi}\int_{4\mpid}^{\tm} dt'\, {\Delta_1^1(t')\sin\Phi_1^1(t')
\over {\Omega^1_1}_R(t') t' (t'-t)}
\nonumber \\
&&\phantom{\Delta_0^0(t)+{t \Omega_0^0(t)\over \tm-t}\Bigg[\alpha_0}
+{1\over\pi}\int_\tm^\infty dt'\,{\vert g_1^1(t')\vert \sin \Phi_1^1(t')\over
{\Omega^1_1}_R(t') t' (t'-t)} \Bigg]
\ena
Notice that the integrands are singular when $t'\to t_m$, since
the Omn\`es function behaves as
\be\lbl{omegrlim} 
{\Omega^I_l}_R(t')\sim \vert t'-t_m\vert ^{\phi_l^I(t_m)\over\pi}
\en
but the singularity is integrable. When $t\to t_m$ the integrands diverge
but this is compensated by the factor of $\Omega_l^I(t)$
multiplying the integrals. It can be shown that 
the solution satisfies automatically
the first matching condition (details of the proof are given in 
App.~\ref{sec:matchingg})
\be\lbl{gijlim}
\lim_{\epsilon\to0} g_l^I(t_m\pm \epsilon)\vert_{\rm sol}=
 g_l^I(t_m )\vert_{\rm input}  \ .
\en
Here, $g_0^0(t)$, $g_1^1(t)$ are treated  in a somewhat different way from that
in ref.~\cite{abm}. In that work, an additional subtraction constant was 
introduced and the values of the  subtraction parameters were fixed 
by imposing that the values of $g^1_1(0)$,  $g^0_0(0)$ 
and ${d\over dt}g^0_0(0)$ be equal to the ChPT prediction at order
$p^2$. Now, the behaviour around $t=0$ is entirely determined 
by solving the full set of equations with the appropriate boundary 
conditions -- our constraints are dispersive
and do not rely on ChPT results. 

At this stage, the formulas~\rf{omformules},\rf{omformules1} 
for $g_0^0(t)$, $g_1^1(t)$ involve 
three parameters: the two 
$S$-wave scattering lengths $a_0^{1/2}$, $a_0^{3/2}$ that appear in the
expressions for $\Delta_0^0(t)$, $\Delta_1^1(t)$ and an additional parameter
$\alpha_0$. We will now clarify their role.

\subsection{Matching conditions and uniqueness} \label{sec:unique}

Once $g_0^0(t)$, $g_1^1(t)$ are expressed according to  
eqs.~\rf{omformules},\rf{omformules1},
the set of four RS equations \rf{royfirst} becomes a closed set of equations
for the four $\pi K$ partial 
waves $f^I_l(s)$, $l=0,1$, $I={1\over2},\ {3\over2}$ . The 
structure of these equations is similar to that of $\pi\pi$ Roy equations:
the kernels consist of a singular Cauchy part and a regular part,
and elastic unitarity provides a non-linear relation between $\re f^I_l(s)$ and
$\im f^I_l(s)$. The equations must be solved with the boundary condition
that the solution phase shifts must equate the input phase shifts at
the frontier of the region of resolution (matching condition).
Therefore, we can apply 
the results  derived recently~\cite{gasser-wanders,wanders} 
concerning the number of independent solutions 
in the vicinity of a given solution.
The multiplicity index of one
solution is determined by the values of the input phase shifts at the
matching point $s=s_m$ (with $s_m\simeq 0.935$ GeV$^2$).
The experimental phase shifts at $s=s_m$ lie in the following ranges
\be
0<\delta_0^{1/2}(s_m)<{\pi\over2},\ \ 
{\pi\over2}<\delta_1^{1/2}(s_m)<{\pi },\ \ 
\delta_0^{3/2}(s_m)<0,\ \ 
\delta_1^{3/2}(s_m)<0. 
\en
According to the discussion in ref.~\cite{wanders}, the multiplicity index
in this situation is $m=0+1-1-1=-1$, to be compared with $m=0$ in the case
of $\pi\pi$. This means that our situation corresponds to a constrained 
system: 
a solution will not exist unless the two $S$-wave scattering lengths
lie on a one dimensional curve.

In practice, however, the phase shift for the $I={3\over2}$,
$P$-wave is extremely small below 1 GeV and the experimental input 
is not precise enough to implement matching conditions
in this channel in any meaningful way (see fig.~\fig{royphaseP32} below).
This leads us to treat the $I={3\over2}$ P-wave on the same footing
as the partial waves with $l\ge 2$. For instance, the dispersive
representations can be projected on $l=2$ and 
used to compute the real part of  $f^{1/2}_2(s)$ for $s\le s_m$ while the 
contribution of $\im f^{1/2}_2(s')$ for $s'\le s_m$ in the integrands
is negligibly small
compared to contributions from $S$- and $P$-waves; it  can be 
evaluated approximately or even ignored~\footnote{
A second argument to neglect the low-energy contribution
of the imaginary part of this partial wave is provided by the chiral counting 
$\im f_1^{3/2}=O(\im f^I_{l\ge2}) =O(p^8)$.}.
Dropping one matching condition,
the effective multiplicity index  becomes $m=0$ for $\pi K$. 
The fact that the multiplicity index  vanishes means 
that solutions should exist for arbitrary values of the two $S$-wave 
scattering lengths $a_0^{1/2}$, $a_0^{3/2}$ lying in some 
two dimensional region, and each solution is unique. 

However, not all solutions are physically acceptable.
An acceptable solution must satisfy the further
requirement that it displays no cusp at the matching point~\cite{acgl}. 
This condition leads to constraints on the subtraction parameters.
First, let us consider the $t$-channel, for which we choose the matching point
$t_m$ to be slightly larger than the $K\Kbar$ threshold. 
As discussed in the previous section,
the solution for $g_0^0(t)$ 
involves one parameter $\alpha_0$. While the equality
$g_0^0\vert_{\rm sol}=g_0^0\vert_{\rm input}$ is automatically guaranteed 
by eq.~\rf{omformules}, the solution
$g_0^0\vert_{\rm sol}$ exhibits a sharp cusp at the matching point in general. 
Therefore, the no-cusp condition 
fixes the value of $\alpha_0$. The same reasoning
can be applied to the $\pi K$ partial waves: imposing the no-cusp condition
to the $I={1\over2}$ $S$- and $P$-waves provides two equations 
which should determine, in principle, the two scattering 
lengths $a_0^{1/2}$, $a_0^{3/2}$. In other words, given ideal 
experimental input data\footnote{ The data are assumed to be ideal also
in the sense that they ensure the existence of a solution to the 
equations\cite{wanders}.}
with no errors in the ranges $s\ge s_m$ and 
$t\ge t_m$, one should be able to fix exactly the two scattering lengths
by solving the RS equations with the appropriate boundary conditions
on the values and the derivatives of the phase shifts. Obviously, the actual
situation is different from that ideal view: the input
data are known with errors and only for discrete values of the energy,
which introduces uncertainties on the boundary conditions and thus
on the solutions of the RS equations.
This point will be addressed in the following section.

\section{Numerical solutions and results} \label{sec:results}

\subsection{Numerical determination of the solutions}\label{sec:numerical}

We have described how to solve the RS equations for the $\pi\pi\to
K\bar{K}$ partial waves.
Assuming that the input for $s>s_m$ is given as well as the input for 
$l\ge 2$ at all energies, our  
purpose is to determine the three phase shifts 
\be
\delta^0(s)\equiv \delta_0^{1/2}(s) ,\quad \delta^1(s)
\equiv\delta_1^{1/2}(s) 
\quad \delta^2(s)\equiv\delta_0^{3/2}(s) 
\en
in the range $\mpd \le s\le s_m$, so that the Roy-Steiner equations
represented symbolically as
\be \re f^a(s)\equiv{s\over \lambda_s}\sin(2\delta^a(s))=\Phi^a[ \delta^{b},s]
\en
are satisfied up to a certain accuracy. We introduce a set of $N$ 
mesh points $\mpd <  s_i \le s_m$ ($N$ was varied between 16 and 30, the
results were very stable) and characterize the accuracy of an
approximate solution by the measure
\be
\epsilon= \max_{i,a} \vert \re f^a(s_i)-\Phi^a[ \delta^{b},s_i ]\vert\ .
\en
An exact solution, of course, satisfies $\epsilon=0$. While it is possible to 
search directly for minimums of $\epsilon$, a more appropriate 
quantity for minimization algorithms is the chi-square
\be
\chi^2= \sum_{i=1}^N\sum_{a=0}^2 \vert \re f^a(s_i)- \Phi^a[ \delta^{b},s_i ]\,
\vert^2
\ ,
\en
which we have minimized using the MINUIT package~\cite{minuit}. 
Approximations to the $\pi K$ phase shifts $\delta^a(s)$
are constructed in the form of polynomials
or piecewise polynomial parametrizations (we tried several forms) 
similar to that proposed by Schenk~\cite{schenk}. 
This is essentially  the same 
method as in ref.~\cite{acgl} for the $\pi\pi$ Roy equations. The 
parameters are constrained so that the phase shifts are continuous
at the matching point and the no-cusp condition
applies to $\delta^0(s)$ and $\delta^1(s)$. 
As discussed in sec.~\ref{sec:unique}, 
these additional conditions fix the values of the 
two $S$-wave scattering lengths, which are therefore included as two 
additional parameters in the minimization of the chi-square. 

Let us denote by $n^{(a)}$ the number of parameters in the representation of
$\delta^a$. Taking $n^{(0)}=3$, $n^{(1)}=2$, $n^{(2)}=1$ 
we obtain an approximation
to the equation with $\epsilon\simeq 5\cdot 10^{-3}$. Adding one more parameter
with $n^{(2)}=2$ makes $\epsilon$ go down to $\epsilon\simeq2\cdot 10^{-3}$ and
with still one more parameter, $n^{(2)}=3$, one 
obtains $\epsilon\simeq1\cdot 10^{-3}$. This provides
good evidence that the approximations are converging to a true solution.
Seeking a much higher accuracy would be difficult:
all integrals must be evaluated with a numerical precision better
than $\epsilon$, and the computation of the phase shifts involve
up to three successive numerical integrations 
(see eqs.~\rf{royfirst},\rf{omdef},\rf{omformules},\rf{omformules1}).

\begin{figure}[htb]
\begin{center}
\includegraphics[width=12cm]{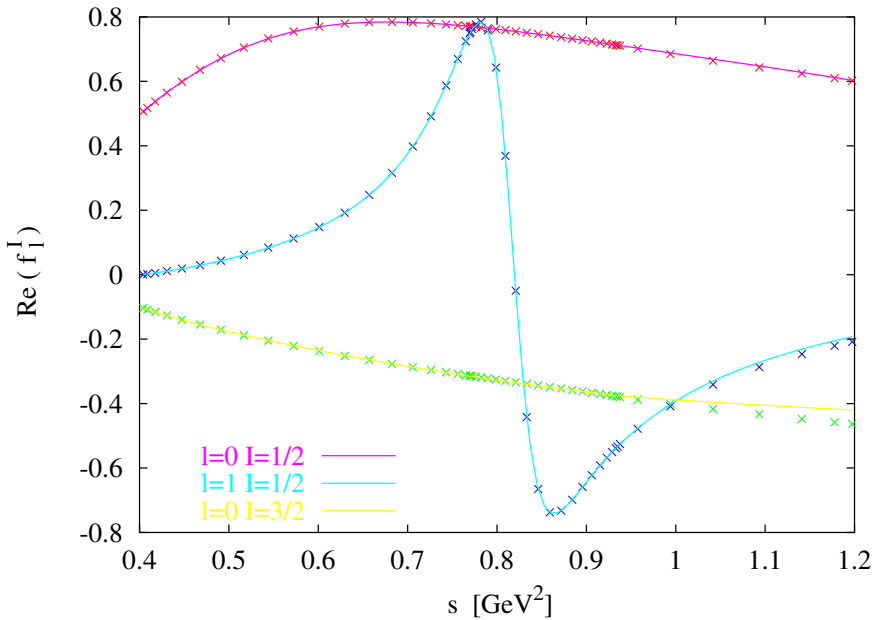}
\caption{\sl Left-hand sides of the RS equations.~\rf{royfirst} (lines) 
compared to the
right-hand sides (points) after minimization in the range $\mpd\le s
\le 0.93$ GeV$^2$ .}
\lblfig{royres}
\end{center}
\end{figure}

The accuracy of the solutions is illustrated  in fig.~\fig{royres}. 
In particular, the figure shows that the 
left- and right-hand sides of the RS equations still agree with a good 
accuracy well above the matching point\footnote{We are then exceeding
the strict domain of applicability of the equations but they are still
expected to be satisfied approximately.}. 
This constitutes  a consistency condition as discussed in ref.~\cite{acgl}. 
We have checked that its fulfilment is a direct consequence of 
imposing the no-cusp conditions. At this level, there
is a notable difference between the $\pi\pi$ and the $\pi K$ RS equations.
In the case of $\pi\pi$ scattering~\cite{acgl}, it is found that 
imposing a {\sl single} no-cusp condition
for the $P$-wave is sufficient to ensure that the 
no-cusp condition holds to a good approximation for the $S$-waves as well, 
and the consistency conditions are well satisfied. 
In the $\pi K$ case, we find that it is necessary to
impose no-cusp conditions for the two phase shifts 
$\delta_0^{1/2}(s)$ and $\delta_1^{1/2}(s)$. 
In fact, even after doing so, we find that a
(small) cusp remains for the third phase shift $\delta_0^{3/2}(s)$.  
This does not represent a serious problem, in practice, because this phase
shift is not determined very precisely in the vicinity of the matching point.
\begin{figure}[hbt]
\begin{center}
\includegraphics[width=16cm]{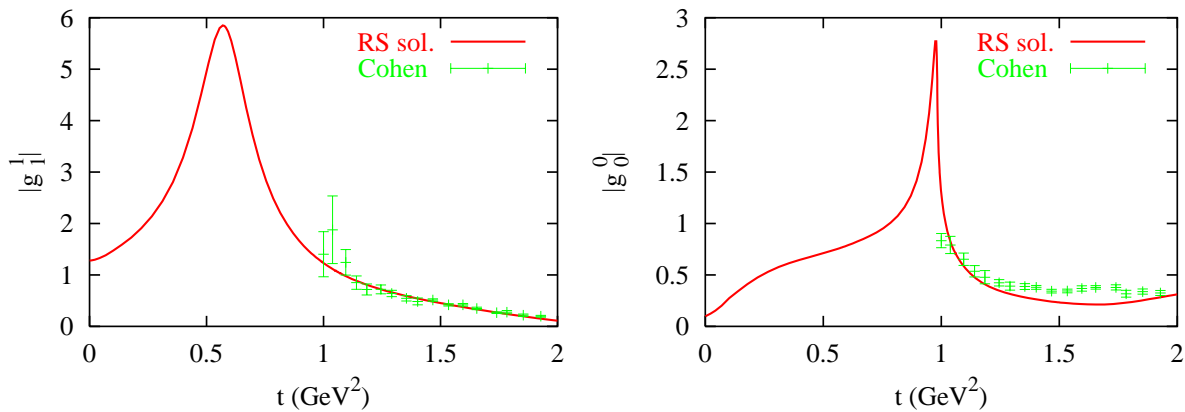}
\caption{\sl Comparison of the absolute values of $g_0^0$ and $g_1^1$
obtained from solving the RS equations
and the corresponding experimental input from ref.~\cite{cohen}.}
\lblfig{g00g11}
\end{center}
\end{figure}

Further consistency conditions ought to be considered in the 
$\pi\pi\to K\Kbar$ sector. Here as well, one expects that the RS equations
should  be approximately satisfied above the matching point. 
This point is illustrated in fig.\fig{g00g11} which compares the 
moduli of $g_0^0(t)$ and $g_1^1(t)$  computed from the RS equations to the
experimental input for these quantities. Very good agreement is observed
for $g_1^1(t)$. In contrast, we find that the agreement for $g_0^0(t)$
is moderately good. In the range $t\ge 4\mkd$ we have checked that
the unitarity bound $\vert S_{\pi\pi\to K\Kbar}\vert\le 1$ 
is  obeyed. 
Adopting a larger value for the matching point $t_m$
improves the input-output agreement for $t>t_m$ but leads to violation
of unitarity for $t<t_m$ close to the $K\Kbar$  threshold.

Another consistency check can be performed. 
In the region where $t\le 4\mpid$, $g_0^0(t)$ and $g_1^1(t)$
can be obtained not only from eqs.~\rf{roylast} which are based on the
fixed$-us$ dispersion relations but also from the fixed$-t$ ones which are
valid in this domain. Both kinds of DR's agree by construction at $t=0$,
the fact that they should continue to agree for negative values of $t$
is not trivial and constitutes a check of consistency of the experimental
input and of the RS solutions. We show these results for $t\le 0$ in 
fig.\fig{gijext}.
\begin{figure}[htb]
\begin{center}
\includegraphics[width=12cm]{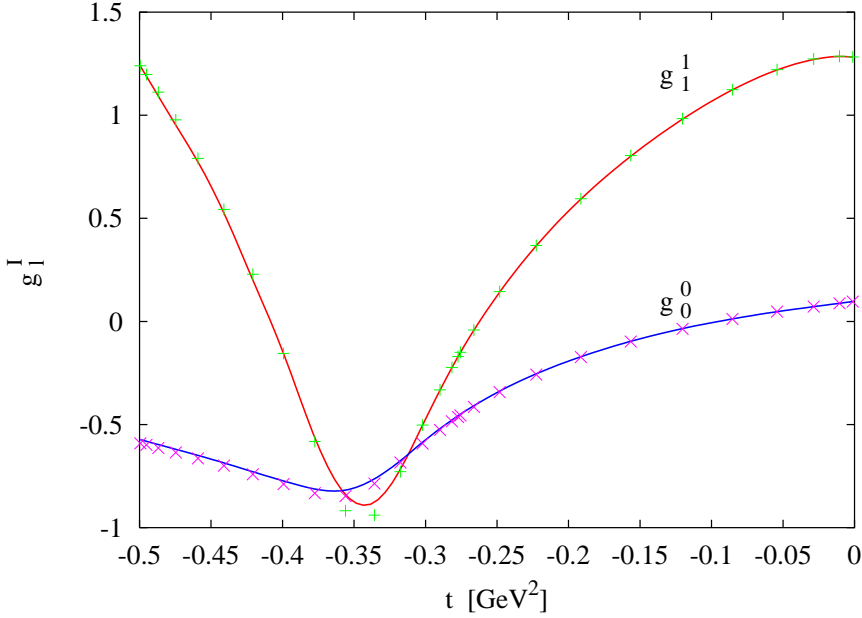}
\caption{\sl Plot of $g_0^0(t)$ and $g_1^1(t)$ in the region $t\le 0$ 
obtained by projecting out two different kinds of dispersion relations.
The lines are obtained from the fixed$-us$ DR's and the points are obtained
from the fixed$-t$ DR's. }
\lblfig{gijext}
\end{center}
\end{figure}

\subsection{Error evaluations and the $S$-wave scattering lengths}

Our general procedure for evaluating the errors consists in performing
variations
of the parameters which enter in the description of the input -- making use
of the errors on these parameters and their covariance matrices as provided
by running the MINUIT package~\cite{minuit}. 
The experimental errors are assumed to be essentially of statistical origin
and the errors at different energy points are assumed to be independent.
Let us discuss first the case of the 
$I={1\over2}$ $S$- and $P$-waves. It is clear that this part of the input 
plays a crucial role as it controls the boundary conditions which determine
the two $S$-wave scattering lengths. To begin with, one notes that variations
of the input in the energy region $E\gapprox 1.5$ GeV has a negligibly small
influence, so we will consider only the energy region 
$\sqrt{s_m}\le E \le 1.5$
GeV. We have performed two different kinds of fits in order to check the
validity of the determination of 
the phase shifts, their derivatives, and the errors obtained from varying
the parameters at the matching point $E=\sqrt{s_m}$. 
Firstly, we perform ``global'' fits based on a K-matrix 
parametrization with six parameters for the $S$-wave and seven parameters
for the $P$-wave. These parameters are determined such as to minimize the
chi-square in the energy region $0.90\le E\le 1.50$ GeV. Secondly, we have
performed ``local'' fits in which one considers separately a small energy
region surrounding the matching point $0.90\le E\le 1.1$ GeV and the remaining
energy region. In the small region we approximate the $S$-wave phase shift
by a quadratic polynomial,
\be
\delta_0^{1/2}(s)=a+b(s-s_m)+c(s-s_m)^2,\quad   0.90\le E\le 1.1\ {\rm GeV}
\en
while for the $P$-wave we use a linear approximation after subtracting the
tail of the $K^*$ resonance
\be
\delta_1^{1/2}(s)-\arctan{m_{K^*} \Gamma_{K^*}\over m^2_{K^*}-s}    =
a+b(s-s_m)\ .
\en
The results from these two fits concerning the input at the matching point
are shown in table \Table{matchpt}. 
One observes that the determinations of the phases
at the matching point are in good agreement as well as that of the errors.
The determinations of the derivative of the $P$-wave agree while those 
of the derivative of the $S$-wave are only in marginal agreement. In this 
case, we consider the determination from the global fit to be somewhat more
reliable as it has continuity and smoothness built in.
\begin{table}[hbt]
\begin{center}
\begin{tabular}{|c|c|c||c|c|}\hline\hline
\ &phase & error & derivative & error \\ \hline
$l=0$ global& 46.5   & 0.6& 44.1& 5.8 \\
$l=0$ local & 46.2   & 0.6& 56.9& 6.6\\ \hline
$l=1$ global& 155.8 &  0.4& 148.0&2.8 \\
$l=1$ local & 156.2 &  0.3& 147.4&2.9 \\ \hline\hline
\end{tabular}
\caption{\sl $S$- and $P$-waves inputs at the matching point as determined
from two different types of fit to the data of Aston {\it et al.}~\cite{aston} 
(see text). Phases are in degrees and their derivatives 
in degrees$\times$GeV$^{-1}$. }
\lbltab{matchpt}
\end{center}
\end{table}

We can now derive the constraints on the $S$-wave 
scattering lengths which arise upon solving the RS equations making
use of the available experimental input above the matching point. Let us
first quote some results concerning the errors. Table~\Table{erreurspart} 
shows 
how the errors affecting the various pieces of input propagate to the
two $S$-wave scattering lengths. One can see that the two  main sources
of uncertainty are \emph{a)} the $\pi K$ $I={1\over2}$ $S$-wave and
\emph{b)} the $\pi\pi\to K\Kbar$ $I=0$ $S$-wave. 
In contrast, the influence of the  partial waves with $l\ge2 $ (in which
the Regge region is also included) is rather modest. 
Finally, this analysis generates
the following results for the scattering lengths $a^I_0$,
\be\lbl{elcenter}
m_\pi\,a^{1/2}_0\simeq 0.224 \pm 0.022,\quad 
m_\pi\,a^{3/2}_0\simeq (-0.448\pm 0.077 )\> 10^{-1}\ .
\en
There is a significant correlation between these two quantities, the
correlation parameter is positive and its value is
\be\lbl{corr13}
\rho_{{1\over2}{3\over2}}=0.908\ .
\en
The one-sigma error ellipse corresponding to the above results for
the $S$-wave scattering lengths is represented in fig.\fig{ellipse}. 
Our results are compatible with the band obtained for
$a_0^{1/2},\ a_0^{3/2}$ in ref.~\cite{johannesson}.
We find a much smaller allowed region 
for the scattering lengths simply because we have used considerably better
experimental input for the $S$- and $P$-waves: in the work of 
ref.~\cite{johannesson} no data at all were available for $E\ge 1.1$ GeV.
Predictions from ChPT at $O(p^4)$ for the $S$-wave scattering lengths
were presented in ref.~\cite{bkm2}. They are recalled below
\be\lbl{swavp4}
m_\pi a_0^{1/2}=0.19\pm 0.02\quad  m_\pi a_0^{3/2}=-0.05\pm0.02
\qquad{\rm (ref.\ \cite{bkm2})}
\en

\begin{table}[hbt]
\begin{center}
\begin{tabular}{|c|c|c|c|c|c|c|}\hline\hline
\ &$f^{1/2}_0$  &$f^{1/2}_1$ & $f^{3/2}_0$ & $g_0^0$ & $g_1^1$ & $l\ge 2$ 
\\ \hline 
$10^2 \Delta{a^{1/2}_0}$&
1.89 & 0.28 & 0.40 & 0.79 & 0.05 & 0.23 \\
$10^2 \Delta{a^{3/2}_0}$&
0.55 & 0.09 & 0.39 & 0.32 & 0.14 & 0.11 \\
$10^2 \Delta(a^{1/2}_0-a^{3/2}_0)$&
1.35 & 0.18 & 0.10 & 0.55 & 0.10 & 0.15 \\ \hline\hline
\end{tabular}
\caption{\sl Sources of error arising from different parts of the input
and the resulting errors in the 
determination of the $l=0$ scattering lengths in units of $m^{-1}_\pi$. }
\lbltab{erreurspart}
\end{center}
\end{table}
Within the errors these results appear compatible with those from
the RS equations. A more refined comparison, however, should take the 
correlation into account. Computing the correlation parameter under the same
assumptions as used  in ref.~\cite{bkm2} for the evaluation of the errors
one obtains the standard error ellipse shown in fig.~\fig{ellipse}. One 
observes that the ChPT ellipse is very narrow and does not intersect
the corresponding error ellipse resulting from the RS equations
\footnote{This particular shape reflects two features of the 
scattering lengths $a_0^-$ and $a_0^+$ in the chiral expansion 
at order $p^4$:  
a) they are essentially uncorrelated (the correlation
parameter is $\rho_{-+}\simeq-0.15$), 
b) the error on $a_0^-$ is very small because
it involves a single chiral coupling ($L_5$) which is multiplied by $m^4_\pi$
while $a_0^+$ involves seven chiral parameters which are multiplied by
$m^2_\pi m^2_K$.}. If one judges from the size of the $O(p^4)$ corrections
as compared to the current algebra result, it seems not unreasonable to
attribute the remaining discrepancy to $O(p^6)$ effects.
We quote also our results for the two combinations
of scattering lengths proportional to $a_0^-$, $a_0^+$
\be
m_\pi\,(a^{1/2}_0- a^{3/2}_0)\simeq 0.269\pm 0.015,\quad
m_\pi\,(a^{1/2}_0+2a^{3/2}_0)\simeq 0.134\pm 0.037\ 
\en
which are of interest in connection with the $\pi K$ atom: the square of 
the first combination is proportional to the inverse lifetime of the atom 
and the sum of the two combinations is proportional
to the energy shift of the lowest 
atomic level~\cite{deser}. The correlation parameter for $a_0^-$, $a_0^+$ 
is also positive and its value is
\be
\rho_{-+}=0.925\ .
\en
For comparison, let us mention the results for the combinations 
proportional to $a_0^-$, $a_0^+$ in ChPT,
\be\lbl{p4apm}
m_\pi\,(a^{1/2}_0- a^{3/2}_0)\simeq 0.238\pm 0.002,\quad
m_\pi\,(a^{1/2}_0+2a^{3/2}_0)\simeq 0.097\pm 0.047\quad\quad 
[{\rm ChPT\ {\it O}(p^4)}]\ . 
\en
The uncertainty affecting $a_0^-$ is remarkably small. This, however, 
could be an artifact of the $O(p^4)$ approximation. It remains
to investigate how $O(p^6)$ corrections affect this result.
\begin{figure}[hbt]
\begin{center}
\includegraphics[width=12cm]{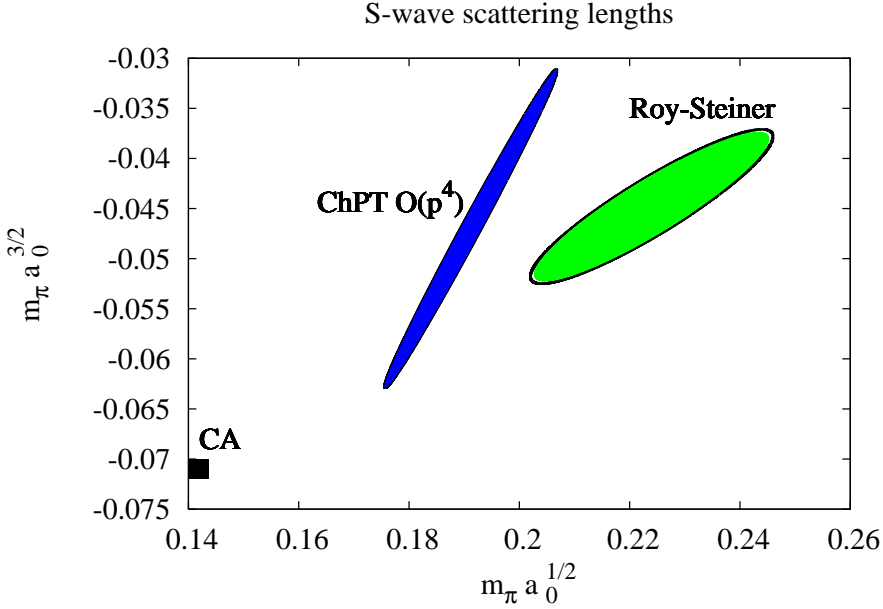}
\caption{\sl Standard error ellipse
for the $S$-wave scattering lengths obtained 
from solving the RS equations with boundary conditions. 
The corresponding ellipse in the ChPT calculation
at $O(p^4)$ and the current-algebra result are also plotted.}
\lblfig{ellipse}
\end{center}
\end{figure}

As discussed above, the fact that the two $S$-wave scattering lengths 
are determined independently (to some extent) comes from imposing the 
no-cusp matching conditions. 
The difference of the two scattering lengths,
$a^{1/2}_0-a^{3/2}_0$, can be determined in an alternative
way from a sum rule~\cite{karabarbounis} (see eq.~\rf{slowsr}).
Using this sum rule one finds 
\be
m_\pi(a^{1/2}_0-a^{3/2}_0)= 0.251 \pm 0.014 \quad\quad [{\rm sum\ rule} ]\ .
\en
In the evaluation, we use our results for the RS solutions in the integration
regions $s'\le s_m$, $t'\le t_m$. The propagation of the experimental 
errors is studied in the same way as explained above. A rather small error is
found, but one must keep in mind that 
the dependence on the asymptotic region is significant here
and it is difficult to evaluate  the error from this region in a very
reliable way.
The central value arising from the sum rule  is smaller than what is
obtained from the RS solution, but the two results  are compatible
within their errors. We also note that the output of the sum rule
is significantly influenced by the values of the scattering lengths
used as input in the integrand. For this reason, the result obtained 
here differs from the one quoted in ref.~\cite{abm}.

\begin{figure}[hbt]
\begin{center}
\includegraphics[width=12cm]{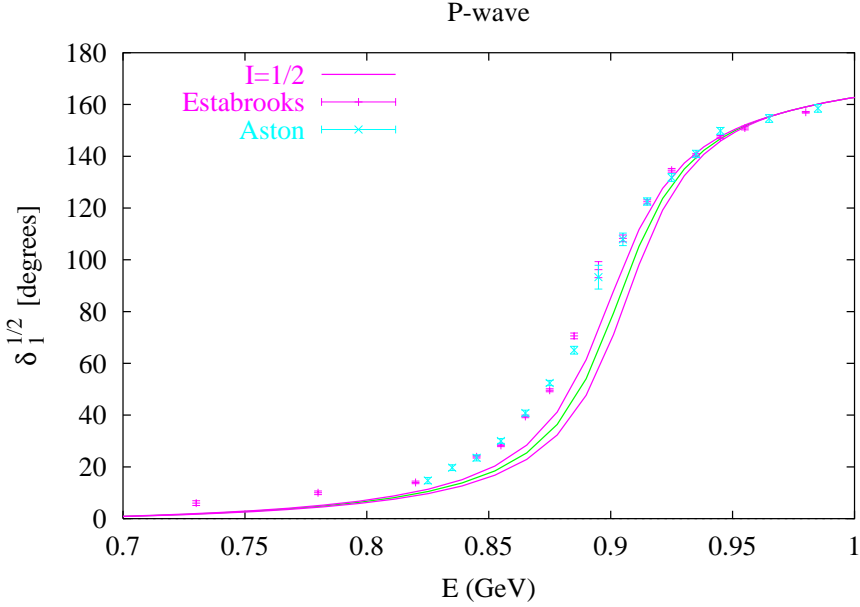}
\caption{\sl The $I={1\over2}$ $P$-wave phase shift obtained from solving
the RS equations. The central curve 
corresponds to solving with $a_0^{1/2},\ a_0^{3/2}$ taken at the center of
the ellipse fig.~\fig{ellipse}. The upper (lower) curves are obtained
by using the points with the maximal (minimal)
values for $a_0^{1/2}$ on this ellipse.}
\lblfig{royphaseP12}
\end{center}
\end{figure}

Before we present the results for the amplitudes in the
threshold region, a few remarks are in order concerning the intermediate 
energy region, that ranges from the threshold up to the matching point. 
Experimental data from production experiments exist below 1 GeV, 
but one has to keep in mind the
possibility that systematic errors may have been underestimated in this
energy region in such experiments 
(a discussion of the $\pi\pi$ case can be found in ref.~\cite{ochs}). 
Fig.~\fig{royphaseP12} shows the $I={1\over2}$ $P$-wave phase shift
from the RS equations compared to experiment. The central curve 
correspond to solving with $a_0^{1/2},\ a_0^{3/2}$ taken at the center of
the ellipse fig.\fig{ellipse} while the upper and lower curves are obtained
by using  the points on the ellipse with the maximal and the minimal
values for $a_0^{1/2}$ respectively. The experimental results 
are seen to deviate from the solutions as the energy
decreases from  the matching point.
In particular, the mass of the $K^*$ which is predicted from
the RS equations is
\be
m_{K^*}=(905 \pm 2)  \ {\rm MeV}
\en
(where $m_{K^*}$ is defined such that $\delta^{1/2}_1(m_{K^*})=\pi/2$) is
nearly 10 MeV larger than the mass quoted in ref.~\cite{aston} 
($m_{K^{*0}}=896 \pm 0.7$ MeV). 
This discrepancy may appear worrying at first sight. It is caused,
in part, by isospin breaking which is not taken into account by 
the RS equations. This could generate an uncertainty of a few MeV as to 
the  value of the
$K^*$ mass that should come out from solving the equations\footnote{For
instance, the result depends on the input values for $m_\pi$ and $m_K$
for which we used $m_\pi=0.13957$ GeV, $m_K=0.4957$ GeV.}.
Besides, it cannot be excluded that the mass of the $K^*$ may not be 
as accurately known as one might believe. 
The determinations of the $K^{*+},\ K^{*0}$ masses used by the PDG
are all based on hadronic production experiments. Recently, a measurement
of the $K^{*+}$ mass based on the $\tau$ decay mode $\tau\to K_S\pi\nu_\tau$
indicated of shift by $4-5$ MeV as compared to the PDG value~\cite{urheim}.
In principle, this method is  more reliable because it is free of any
final state interaction problem, but better statistics are needed to
clarify this issue.

The two $S$-wave  phase shifts predicted by the RS equations are shown in 
fig.~\fig{royphaseS}. For the isospin $I={1\over2}$ the RS solution
does not exhibit any of the oscillations appearing in the 
data of  ref.~\cite{estabrooks}.
For the isospin $I={3\over2}$ phase shift, the experimental data for 
$E<0.9$ GeV lie systematically below the RS curve, 
by 2-3 standard deviations.
The RS equations  also predict the $I={3\over2}$ $P$-wave
phase shift, the result is shown in fig.\fig{royphaseP32}. This 
phase shift displays the unusual feature that it is positive at very low
energy and changes sign as the energy increases. In the region around 1 GeV
the results are in qualitative agreement with the experimental data of
Estabrooks {\it et al.}

\begin{figure}[hbt]
\begin{center}
\includegraphics[width=14cm]{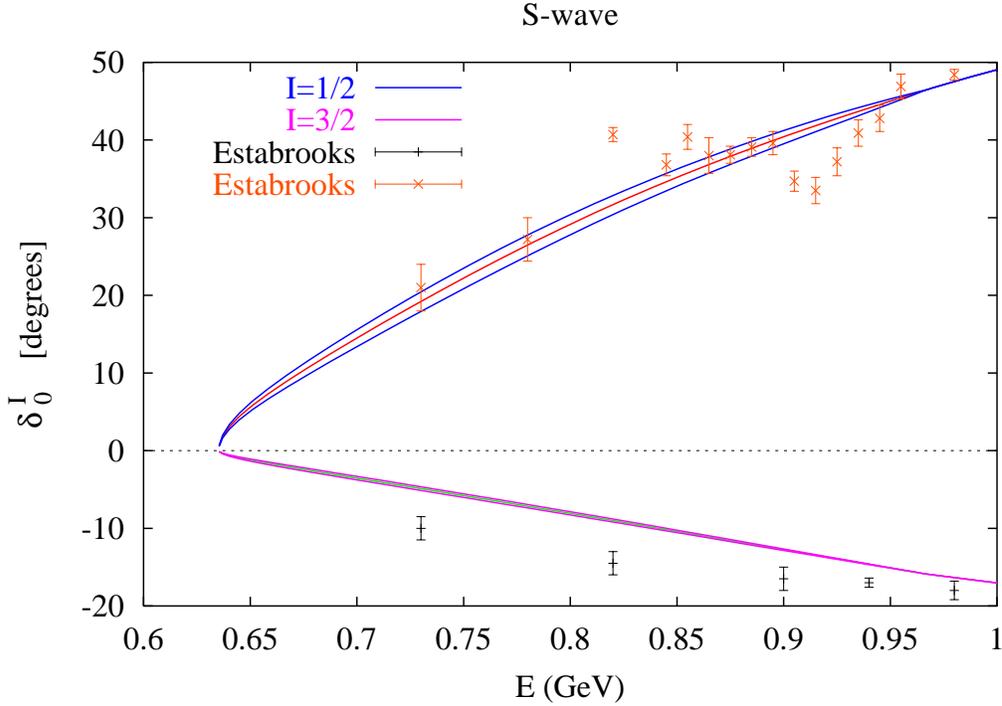}
\caption{\sl Same as fig.~\fig{royphaseP12} for the $I={1\over2}$ $S$-wave
phase shift (curves in the upper half of the figure) 
and the $I={3\over2}$ $S$-wave phase shift (curves in lower half).
}
\lblfig{royphaseS}
\end{center}
\end{figure}

\begin{figure}[hbt]
\begin{center}
\includegraphics[width=12cm]{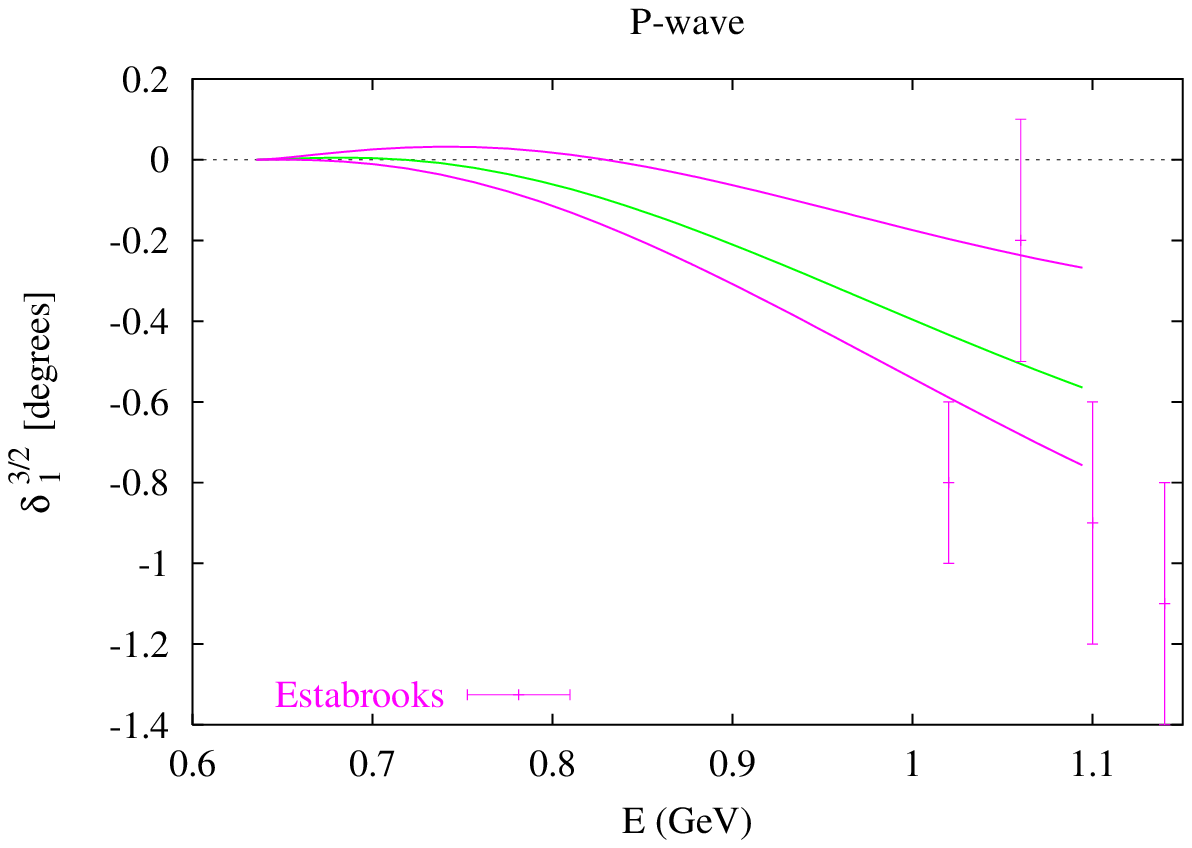}
\caption{\sl Same as fig.~\fig{royphaseP12} for the $I={3\over2}$ $P$-wave
phase shift.}
\lblfig{royphaseP32}
\end{center}
\end{figure}

\subsection{Results for threshold and sub-threshold expansion parameters}

The behaviour of amplitudes at very small energies is conveniently 
characterized by sets of expansion parameters, which are particularly  
useful for making comparisons with chiral expansions. 
We consider first
the set obtained by performing an expansion around the $\pi K$ threshold.
These parameters are conventionally  defined from the partial-wave amplitudes
as follows
\be\lbl{threshold}
{2\over\sqrt s} {\rm Re} f_l^I(s)= 
q^{2l}\left( a_l^I + b_l^I q^2 +c_l^I q^4+\ldots
\right)
\en
with 
\be
s=\mpd + {\mpd q^2\over \mpi\mk} -{\mpd\mmd q^4\over 4\mpi^3\mk^3}+\ldots
\en
Once a solution of the RS equations is obtained, all the threshold parameters
are predicted. The two $S$-wave scattering lengths are determined from
the matching conditions, as explained above. 
The other threshold parameters
may be  obtained from the dispersive representation eq.~\rf{fixtfin}
in the form of sum rules. These are obtained by projecting 
the DR's over the relevant
partial wave and then expanding the variable $s$ in powers of $q^2$. 
Divergences may appear in this process because derivatives are discontinuous
at threshold and it must be specified that the limit is to be taken
from above. This problem is easily handled by computing
some pieces of the integrals analytically as explained in ref.~\cite{acgl}.
The sum rules are evaluated by using RS solutions
below the matching points and the fits to the experimental data above. 
For $l=0,\ 1$ we have computed the parameters $a_l$, $b_l$ and $c_l$ 
in an alternative manner by using our solution for $\re f_l^I(s)$ for
three values of $s$ and solving a linear system of equations. The two methods
were in very good agreement and the
results for the threshold parameters 
are summarized in table \Table{threshpar}. 
The values of the $P$-waves scattering lengths in ChPT at NLO was given
in ref.\cite{bkm2}
\be
m^3_\pi a_1^{1/2}=0.016\pm 0.003\quad m^3_\pi a_1^{3/2}=(1.13\pm 0.57)\,
10^{-3} \qquad{\rm (ref.\ \cite{bkm2})} \ .
\en
Within the errors, these values are compatible with our corresponding results
displayed in table~\Table{threshpar}.
\begin{table}[ht]
\begin{center}
\begin{tabular}{|c|c|c|c|}\hline\hline
\ & $I={1\over2}$             & $I={3\over2}$ &
$\left(I={1\over2}\right)-\left(I={3\over2}\right)$        \\ \hline
$m^3_\pi\>a_1^I$  & $(0.19\pm0.01 )10^{-1}$&$(0.65 \pm0.44 )10^{-3}$ & 
$(0.18  \pm0.01 )10^{-1} $\\
$m^5_\pi\>a_2^I$  & $(0.47\pm0.03 )10^{-3}$&$(-0.11\pm0.27 )10^{-4}$&
$(0.48  \pm0.01  )10^{-3} $\\
$m^7_\pi\>a_3^I$  & $(0.23\pm0.03)10^{-4}$&$(0.91\pm0.30)10^{-5}$ 
&$(0.14 \pm0.01)10^{-4} $
\\ \hline
$m^3_\pi\>b_0^I$  & $(0.85\pm0.04)10^{-1}$&$(-0.37\pm0.03)10^{-1}$ &
$ 0.12 \pm0.01 $\\
$m^5_\pi\>b_1^I$  & $(0.18\pm0.02)10^{-2}$ &$(-0.92\pm0.17)10^{-3}$ & 
$(0.27 \pm0.01)10^{-2} $  \\
$m^7_\pi\>b_2^I$  & $(-0.14\pm0.03)10^{-3}$ &$(-0.96\pm0.26)10^{-4}$& 
$(-0.48 \pm0.02)10^{-4} $
\\ \hline
$m^5_\pi\>c_0^I$  & $(-0.45\pm0.04)10^{-1}$ &$(0.18\pm0.02)10^{-1}$ & 
$(-0.62\pm0.06)10^{-1} $   \\
$m^7_\pi\>c_1^I$  & $(0.71\pm0.11)10^{-3}$ & $(0.51\pm0.09)10^{-3}$& 
$(0.20\pm0.03)10^{-3} $
\\ \hline\hline
\end{tabular}
\caption{\sl Results for the threshold expansion parameters 
(see eq.~\rf{threshold} ) computed 
from the dispersive representations and the RS
equations solutions. The third column  shows the difference
of the $I={1\over2}$ and the $I={3\over2}$ parameters}
\lbltab{threshpar}
\end{center}
\end{table}

ChPT expansions of the amplitude are expected to have 
best convergence properties
in unphysical regions away from any threshold singularity.
The dispersive representations derived in sec.~\ref{sec:ampldisp}
allow us to evaluate the amplitude in such regions. A first domain
considered in the literature is the neighbourhood of the 
point $s=u$, $t=0$. The
following set of expansion parameters are conventionally introduced
\bea\lbl{subthreshold}
&& F^+(s,t)=\sum \tilde C^+_{ij}\, t^i \nu ^{2j}\nonumber\\
&& F^-(s,t)=\nu\, \left\{\, \sum \tilde C^-_{ij}\, t^i \nu ^{2j}\right\}
\ena
where
\be
\nu={s-u\over4m_K}\ .
\en 
It is customary to quote the values of the dimensionless parameters
$C^\pm_{ij}$ which are related to $\tilde C^\pm_{ij}$ by
\be\lbl{subthreshold1}
C^+_{ij}= (m_{\pi^+})^{2i+2j}\, \tilde C^+_{ij},\quad
C^-_{ij}= (m_{\pi^+})^{2i+2j+1}\, \tilde C^-_{ij}\ .
\en
The results for the subthreshold expansion parameters are collected in 
table \Table{subthreshpar}. The table also shows for comparison
results from ref.~\cite{lang-porod}, which used fits to the
experimental data of Estabrooks {\it et al.}~\cite{estabrooks} combined
with earlier sets of data (taking into account the data points 
below 1 GeV as well as above). The authors of ref.~\cite{lang-porod} 
observed that the low-energy part of the data of Estabrooks {\sl et al} 
leads to  inconsistencies with a dispersive representation of $F^-$.
The agreement with our results is
reasonable  for the coefficients $C^-_{i j}$. For the coefficients
$C^+_{i j}$ the results are compatible within the errors, except
for the coefficient $C^+_{1 0}$, for which we find a somewhat larger
value.
Another point of interest is the Cheng-Dashen point $s=u$, $t=2m^2_\pi$.
The value of the amplitude $F^+$ at this point can be related to the
kaon sigma-term~\cite{chengda} (see~\cite{gasser-sainio} for a recent review).
We obtain for this quantity
\be
F^+(\mkd,2\mpid)=3.90 \pm 1.50\ .
\en  

\begin{table}[hbt]
\begin{center}
\begin{tabular}{|c|c||c|c|c|c|c|}\hline\hline
          &           &             &               &ref.~\cite{lang-porod}
&$SU(3)$&$SU(2)$\\\hline
          &           &  $C^-_{00}$ &$8.92\pm0.38$&$7.31\pm0.90$ &2&1\\
\multispan7\dotfill\\
$C^+_{00}$& $2.01\pm1.10$&              &          &$-0.52\pm2.03$&2&2\\
\multispan7\dotfill\\
$C^+_{10}$& $0.87\pm0.08$&             &          &$0.55\pm0.07$ &2&2\\ 
\hline
$C^+_{01}$& $2.07\pm0.10$         &            &  &$2.06\pm0.22$  &4&2\\
\multispan7\dotfill\\
          &           & $C^-_{10}$ &$0.31\pm0.01$ &$0.21\pm0.04$  &4&3\\
\multispan7\dotfill\\
$C^+_{20}$& $(0.24\pm 0.06)10^{-1}$&           &  &               &4&4 \\
\hline
          &           & $C^-_{01}$&$0.62\pm0.06$   &$0.51\pm0.10$&6&3\\
\multispan7\dotfill\\

$C^+_{11}$&$(-0.66\pm0.10)10^{-1}$&           &    &$-0.04\pm0.02$&6&4\\
\multispan7\dotfill\\
          &           &$C^-_{20}$ &$(0.85\pm0.01)10^{-2}$&        &6&5\\
\multispan7\dotfill\\
$C^+_{30}$& $(0.34\pm0.08)10^{-2}$&           &   &                &6&6 \\
\multispan7\dotfill\\
$C^+_{02}$& $0.34\pm0.03         $&           &   &                &8&4 \\
\hline\hline
\end{tabular}
\caption{\sl  Results for the dimensionless subthreshold expansion 
parameters defined in eqs.~\rf{subthreshold} and \rf{subthreshold1}. 
The last two columns indicate the chiral order of the
leading tree-level contribution to each parameter in $SU(3)$ 
and $SU(2)$ ChPT respectively.}
\lbltab{subthreshpar}
\end{center}
\end{table}

\subsection{Some implications for the $O(p^4)$ chiral couplings}
In this section we  present some results on the $O(p^4)$ couplings
of the $SU(3)$ chiral expansion, which are easily derived from
the subthreshold parameters obtained above. More detailed
comparisons between chiral expansions and dispersive representations
of the $\pi K$ scattering amplitude should be performed, but this is
left for future work. The expression for this amplitude in ChPT at
order $p^4$ was presented in ref.~\cite{bkm1}. More specifically,  
we will make use 
of a reformulation of the expression of ref.~\cite{bkm1} in which $F_0$
is expressed in terms of $F_\pi$ only (and not $F_K$) as in ref.~\cite{abm}
(a factor $\bar J_{\pi K}(s)$ is missing in eq. (41)  of that reference).
From this, it is straightforward to obtain the chiral formulas 
for the subthreshold expansion parameters. We present these in numerical
form below, in which we use the following values for the masses, 
the pion decay constant and the renormalization scale $\mu$ 
\be
m_\pi=0.13957,\ m_K=0.4957,\ m_\eta=0.5473,\ F_\pi=0.0924,\ \mu=0.77\ 
\rm {(all\ in\ GeV)}\ .
\en
The dimensionless subthreshold parameters $C^+_{ij}$ then have the following 
numerical expressions in ChPT at NLO 
\bea
&& C^+_{00}=0.14985 +{8\mpid\mkd\over F^4_\pi}\left[ 4L_1+L_3-(4L_4+L_5)
+2(L_8+2L_6)\right]\nonumber\\
&& C^+_{10}=0.45754 +{4(\mkd+\mpid)\mpid\over F^4_\pi}\left[
-(4L_1+L_3)+2L_4\right]
+{2 m^4_\pi \over F^4_\pi} L_5\nonumber\\
&& C^+_{20}=0.02554+{2 m^4_\pi \over F^4_\pi}\left[4L_1+L_2+{5\over4}L_3
\right] \nonumber\\
&& C^+_{01}=1.67285+{8\mkd\mpid\over F^4_\pi}\left[4L_2+L_3\right]
\ena
while the subthreshold parameters $C^-_{ij}$ read
\bea
&& C^-_{00}=8.42568 +{8 m_K m^3_\pi\over F^4_\pi} L_5\nonumber\\
&& C^-_{10}=-0.02533-{4m_K m^3_\pi\over F^4_\pi} L_3 \ .
\ena
In order to lighten the notations we have denoted the renormalized couplings 
$L^r_i(\mu=0.77)$ simply by $L_i$. It is now easy to solve for the $L_i$'s
making use of the results from table \Table{subthreshpar}, the results 
for $L_1,...,L_4$ are
collected in table \Table{Li}. The errors are obtained, as before,
by varying all the parameters of the fits to the input data and taking
into account the correlations. These errors
appear to be rather small but they only reflect  the uncertainty coming
from the input data. 
The dominant  source of uncertainty in the determination of the $L_i$'s 
comes from the unknown higher-order terms in the chiral expansion,
this uncertainty is expected to be of the order of 30-40\% . 
This can be seen from the table which shows
the results of some alternative determinations based on the 
$Kl_4$ form factors~\cite{riggenbach,bijnens,amoros} and on $\pi K$ 
sum rules~\cite{abm}\footnote{In that paper, terms of order $p^6$ were 
dropped in the dispersive representations and the phase shifts used
below 1 GeV in the sum rules were not constrained to obey the RS equations.}. 
We also quote the results that we get for $L_5$
and for $L_8+2L_6$ which have rather large errors
\be
10^3 L_5= 3.19\pm 2.40\quad    10^3 (L_8+2L_6)= 3.66\pm 1.52\ .
\en
The coupling $L_5$  is determined, in principle,  
from $C^-_{00}$ but its contribution turns out to be suppressed, as it appears
multiplied by a factor $m^2_\pi$.
In order to determine $L_8+2L_6$ we used
$C^+_{00}$ and the value $L_5\simeq 1.4\cdot10^{-3}$ derived 
from $F_K/F_\pi$.
The large uncertainty for $L_8+2L_6$ reflects 
that affecting the coefficient $C^+_{00}$ or, alternatively,
the uncertainty in  the combination of scattering lengths 
$a^{1/2}_0+2a^{3/2}_0$. This could improve considerably once experimental
results from $\pi K$ atoms are available. Our result for $L_4$, though affected
by a sizeable error, agrees with the evaluations~\cite{moussall,moussall0}
based on a dispersive method for constructing scalar form factors~\cite{dgl}
and is suggestive of a significant violation of the OZI rule in the scalar
sector.
\begin{table}[hbt]
\begin{center}
\begin{tabular}{|c||c|c|c|c|}\hline
   &$\pi K$ Roy-Steiner & $\pi K$ sum-rules & $Kl_4,\ O(p^4)$ &  
$Kl_4,\ O(p^6)$ \\ \hline
$10^3\, L_1$  & $1.05\pm 0.12$ &  $0.84\pm 0.15$ & $0.46\pm 0.24$ & $0.53\pm 0.25$ \\
$10^3\, L_2$  & $1.32\pm 0.03$ &  $1.36\pm 0.13$ & $1.49\pm 0.23$ & $0.71\pm 0.27$ \\
$10^3\, L_3$  & $-4.53\pm0.14$ & $-3.65\pm 0.45$ & $-3.18\pm0.85$ & $-2.72\pm 1.12$\\
$10^3\, L_4$  & $0.53\pm 0.39$ & $0.22\pm 0.30 $ & $ $   & $-0.2\pm 0.9$ \\
\hline
\end{tabular}
\caption{\sl Chiral couplings $L_i^r(\mu)$, $\mu=0.77$ GeV obtained
by matching the dispersive results for  the subthreshold expansion
parameters (see table~\Table{subthreshpar}) with their chiral expansion
at order $p^4$.
Also shown are the results from ref.~\cite{abm} (col. 3) as
well as those from ref.~\cite{amoros} in which fits to the $Kl_4$ form factors
were perfomed using chiral expansions at order $p^4$ (col. 4) 
as well as  $p^6$ (col. 5).
}
\lbltab{Li}
\end{center}
\end{table}

\section{Conclusions}

In this paper, we have set up and then solved a system of equations \`a la
Roy and Steiner for the $S$- and $P$-partial waves of the $\pi K\to \pi K$ 
and the $\pi\pi\to K\Kbar$ amplitudes. These
equations are necessary consequences of analyticity and crossing, together with
plausible assumptions concerning the range of effective 
applicability of elastic unitarity. 
In treating these equations, the approach advocated recently 
in ref.~\cite{acgl}
was followed, which consists in choosing a matching point around 1 GeV 
and enforcing a set of boundary conditions at this point.
As  input for this analysis, we have 
exploited for the first time the high-statistics data which are now
available from $KN\to K\pi N$ as well as $\pi N\to K\Kbar N$ production
experiments. 

The main result obtained from solving the RS equations together with the
boundary conditions is the determination of an allowed region for the two
$S$-wave scattering lengths which is shown, as a one-sigma ellipse,
in fig.~\fig{ellipse}. This region is much smaller than the ones resulting from
older analyses, e.g. ref.~\cite{johannesson}; this simply reflects the
better accuracy of the experimental input data used here.
Using this result together with  the dispersive representations one 
can determine the $\pi K$ scattering amplitude in regions of very low 
energies inaccessible to experiment. We have computed a set of sub-threshold
expansion parameters and then matched the result with the $SU(3)$ ChPT 
expansion of the amplitude at NLO~\cite{bkm1,bkm2}. This leads to a 
determination of the Gasser-Leutwyler coupling constants $L_1$, $L_2$, $L_3$,
and $L_4$. Comparisons with previous results is suggestive
of significant $O(p^6)$ effects but certainly not so large as to
invalidate the $SU(3)$ expansion.
The bounds that we have obtained for the $S$-wave scattering 
lengths constrain the combination $2L_6+L_8$. 

The value of $L_4$ is of particular interest. Since this low-energy
constant violates the Zweig rule in the scalar channel, its value
is related to the role of sea-quark effects and to the link between
the $SU(2)$ and $SU(3)$ chiral limits~\cite{dgs1,dgs2}. 
Moreover, it was recently pointed out
that the value of $L_4$ can be used to discriminate between different 
assignments for the scalar-meson multiplets~\cite{ecker}, such a connection 
was also illustrated in ref.~\cite{moussall}. The value that we found
is in agreement with the determination based on the scalar 
form-factors~\cite{moussall0,moussall,dhonte} but disagrees with the
prediction from the chiral unitarization model~\cite{pelaez}.
More detailed comparisons with ChPT expansions should be performed
but this is left for future work.
At present, the amplitude has been computed at order $p^4$ in 
the three-flavour expansion and, more recently, in the 
two-flavour expansion~\cite{roessl} (see also~\cite{kubis}). 
The latter is expected to have  better convergence but it is less 
predictive: let us however remark that 
the expression of the antisymmetric amplitude $F^-$
involves only three $SU(2)$ chiral parameters. 

Another topic of interest in connection with $\pi K$ scattering is the 
problem of localizing unambiguously  
a possible $\kappa$ meson (see ref.~\cite{cherry} for a review of
the literature). 
 A naive test based on the collision time concept~\cite{Kelkar} applied 
 to our results for the  $S$-wave phase shift gives no indication 
 for a resonance. 
In principle, our results provide an improved and more
complete input for an analysis such as performed in ref.~\cite{cherry}. 

Dispersive analyses, of course, cannot replace 
low energy measurements. 
Much more stringent constraints on the $S$-wave scattering lengths
could be derived from the RS equations if reliable data were available
at low energy. 
For instance, the analysis could be much improved soon, 
once low-energy data on the $P$-wave phase shifts
are obtained  from the $\tau\to K\pi\nu_\tau$ decay.
In the long term, $S$-wave phase shifts could be measured in 
$D\to K \pi l \nu_l$ decays~\cite{focus}. 
Finally, direct measurements of combinations of  $S$-wave scattering
lengths are planned, based on forming $\pi K$ atoms and measuring their
lifetime and the shift of the lowest atomic level~\cite{katom} (see
refs.~\cite{nehme,kubis2} for a discussion of  related theoretical issues).

\noindent{\bf Acknowledgments: }
We are grateful to J. Stern for his interest, discussions and 
suggestions. B.M. would like to thank B. Ananthanarayan for useful  
remarks and M.R. Robilotta for offering him a copy of H\"ohler's book.
P.B would like to thank the IPN Orsay for its hospitality and
financial support during his stay in Paris.

\appendix

\section{Continuity of  \boldmath{$g^1_1(t)$} at 
\boldmath{$t=t_m$}} \label{sec:matchingg}

In this appendix we prove the validity of eq.~\rf{gijlim}
\be\lbl{conteq}
   \lim_{\epsilon\to 0}g^I_l(t_m\pm\epsilon)|_{\mbox{sol}}
          = g^I_l(t_m)\vert_{\mbox{input}}\nonumber
\en
for $g^1_1(t)$. We will consider the limit from below, $t\to t^-_m$, 
the other limit can be handled in an exactly similar way. 
Let us start from eq.~\rf{omformules1} for $g_1^1(t)$
which expresses the solution in terms of the input values for
the phase $\Phi_1^1(t')$ and the modulus $\vert g_1^1(t')\vert$.
\be
g^1_1(t)|_{\mbox{sol}} = \Delta^1_1(t) + I_1(t) + I_2(t)
\en
with
\bea
&&    I_1(t) =  \Omega^1_1(t)\frac{t}{\pi}\int_{4 m_\pi^2}^{t_m}d\,t'
		    \frac{\Delta^1_1(t')\sin\Phi^1_1(t')}
		    {{\Omega^1_1}_R(t') {t'}(t'-t)},\nonumber\\
&&    I_2(t)  =  \Omega^1_1(t)\frac{t}{\pi}\int_{t_m}^{\infty}d\,t'
		    \frac{|g^1_1(t')|\sin\Phi^1_1(t')}
		    {{\Omega^1_1}_R(t') {t'}(t'-t)},
\ena
which behaviour when $t\to t^-_m$ has to be investigated.
In a first step, one writes $I_{1,2}$ as
\bea\lbl{decoup}
&&   I_1(t)  = \Omega^1_1(t)\frac{t}{\pi}\left\{\int_{t_m-a}^{t_m}d\,t'
		\frac{\Delta^1_1(t')\sin\Phi^1_1(t')}{{\Omega^1_1}_R(t'){t'}(t'-t)}+
		\int_{4 m_\pi^2}^{t_m-a}d\,t'
		\frac{\Delta^1_1(t')\sin\Phi^1_1(t')}{{\Omega^1_1}_R(t'){t'}(t'-t)}
		\right\},\nonumber\\
&&   I_2(t)  = \Omega^1_1(t)\frac{t}{\pi}\left\{\int_{t_m}^{t_m+a}d\,t'
		\frac{|g^1_1(t')|\sin\Phi^1_1(t')}{{\Omega^1_1}_R(t'){t'}(t'-t)}+
		\int_{t_m+a}^{\infty}d\,t'
		\frac{|g^1_1(t')|\sin\Phi^1_1(t')}{{\Omega^1_1}_R(t'){t'}(t'-t)}
		\right\},
\ena
where $a$ is a small positive number. When $t\to t^-_m$ the modulus of
$\Omega_1^1$ goes like
\be
\vert \Omega_1^1(t)\vert\sim\vert t-t_m\vert ^{\Phi_1^1(t_m)\over\pi}
\en
and thererefore vanishes since $\Phi_1^1(t_m)>0$. 
This implies that the second terms of $I_{1,2}(t)$ in eqs.~\rf{decoup}
also vanish when $t\to t^-_m$ because the integrals multiplied  
by $\Omega_1^1$ remain finite.
\bea
&&       I_1(t\to t^-_m)  = \Omega^1_1(t)\frac{t}{\pi}\int_{t_m-a}^{t_m}d\,t'
            \frac{\Delta^1_1(t')\sin\Phi^1_1(t')}{{\Omega^1_1}_R(t'){t'}(t'-t)}
	    \nonumber\\
&&      I_2(t\to t^-_m)  = \Omega^1_1(t)\frac{t}{\pi}\int_{t_m}^{t_m+a}d\,t'
            \frac{|g^1_1(t')|\sin\Phi^1_1(t')}{{\Omega^1_1}_R(t'){t'}(t'-t)}.
\ena
Assuming that $a$ is small enough we can replace ${\Omega^1_1}_R(t')$
by its leading behaviour when $ t'\to t_m$ (eq.~\rf{omegrlim} ) 
and we make the same replacement 
for $\Omega^1_1(t)$. Next, we perform the following change of variables 
in the integrals
\be
t'=(t_m-t) v + t_m\ ,
\en
the limits are then expressed in the following way,
\bea
 &&I_1( t\to t^-_m)=\Delta_1^1(t_m)\exp(i\Phi_1^1(t_m))\sin \Phi_1^1(t_m)
{1\over\pi}\int_0^\infty {dv\over v^\alpha(1-v) }\nonumber\\
&&I_2( t\to t^-_m)= g_1^1(t_m) \sin \Phi_1^1(t_m)
{1\over\pi}\int_0^\infty {dv\over v^\alpha(1+v) }\ .
\ena
with
\be
\alpha={\Phi_1^1(t_m)\over\pi }\ .
\en
The result on the value of $g_1^1(t)|_{\mbox{sol}}$ 
at the matching point follows from
the values of the two definite integrals~\cite{gradstein} 
(which are well defined for $0<\alpha<1$)
\be
{1\over\pi}\int_0^\infty {dv\over v^\alpha(1-v)}=-{\cos(\pi\alpha)\over
\sin(\pi\alpha)} +i,\qquad
{1\over\pi}\int_0^\infty {dv\over v^\alpha(1+v)}={1\over\sin(\pi\alpha)}
\en
as this implies
\be
I_1( t\to t^-_m)=-\Delta_1^1(t_m),\quad I_2( t\to t^-_m)= g_1^1(t_m)\ 
\en
which proves the continuity equation \rf{conteq} for $g_1^1$ 
when $t_m$ is approached from below. Similar arguments can be used to 
prove continuity when  $t_m$ is approached from above. Finally, the proof
is easily generalized to the case of $g_0^0$ which involves one more 
subtraction.

\end{document}